\documentclass[journal]{IEEEtran}
\usepackage{graphicx}
\usepackage{amsmath,amssymb,amsfonts}
\usepackage{booktabs}
\usepackage{tabularx}
\usepackage{textcomp}
\usepackage{url} 
\usepackage{cite}

\usepackage[caption=false,font=footnotesize]{subfig}

\usepackage{algorithm}
\usepackage{algpseudocode}

\usepackage[export]{adjustbox}
\usepackage{makecell}
\usepackage{comment}
\usepackage{mdframed}

\newcounter{formulation}
\newenvironment{formulationN}[1]{%
  \refstepcounter{formulation}%
  \begin{mdframed}[linewidth=0.5pt,roundcorner=5pt,innerleftmargin=4pt,innerrightmargin=4pt]
  \noindent\textbf{Formulation \theformulation.\ #1}\par\vspace{0.5ex}
}{%
  \end{mdframed}
}

\usepackage[hidelinks]{hyperref}

\begin{document}
%
\title{Resource Management and Circuit Scheduling for Distributed Quantum Computing Interconnect Networks}
%
%
%


\author{Sima~Bahrani, Romerson~D.~Oliveira, Juan~Marcelo~Parra-Ullauri, Rui~Wang, and~Dimitra~Simeonidou%
\thanks{This work was supported by Engineering and Physical Sciences Research Council Quantum Communication Hub grant ref. EP/T001011/1, EPSRC Integrated Quantum Networks Hub grant EP/Z533208/1,  International Science Partnerships Fund from Science and Technology Facilities Council (STFC) DyMEND project (grant ref:593) and the Innovate UK-funded project, Quantum Data Centre of the Future (10004793). (\textit{Corresponding authors: Sima~Bahrani and Rui~Wang})}
\thanks{Sima~Bahrani, Rui~Wang, and Dimitra~Simeonidou are with the Smart Internet Lab, School of Electrical, Electronic, and Mechanical Engineering, University of Bristol, Bristol, BS8 1UB, UK (e-mail: sima.bahrani@bristol.ac.uk; rui.wang@bristol.ac.uk; Dimitra.Simeonidou@bristol.ac.uk).}%
\thanks{Romerson~D.~Oliveira was with the Smart Internet Lab, School of Electrical, Electronic, and Mechanical Engineering, University of Bristol, Bristol, BS8 1UB, UK, and is now with Nu Quantum, Cambridge, CB3 0FA, UK (e-mail: romerson.oliveira@nu-quantum.com).}%
\thanks{Juan~Marcelo~Parra-Ullauri was with the Smart Internet Lab, School of Electrical, Electronic, and Mechanical Engineering, University of Bristol, Bristol, BS8 1UB, UK, and is now with BT, Bristol, BS2 0JJ, UK (e-mail: juan.m.parraullauri@bt.com).}%
}
\maketitle

\begin{abstract}
 Distributed quantum computing (DQC) has emerged as a promising approach to overcome the scalability limitations of monolithic quantum processors in terms of computational capability. However, realising the full potential of DQC requires effective resource management and circuit scheduling. This involves efficiently assigning each circuit to a subset of quantum processing units (QPUs), based on factors such as their computational power and connectivity. In heterogeneous DQC networks with arbitrary connectivity topologies and non-identical QPUs, this becomes a complex challenge. This paper addresses resource management and circuit scheduling in such settings, with a focus on computing resource allocation in a quantum data centre. We propose circuit scheduling algorithms based on Mixed-Integer Linear Programming (MILP). Our MILP model accounts for errors arising from inter-QPU communication. In particular, the proposed schemes consider key factors, including network topology, QPU capacities, and quantum circuit structure, to make efficient scheduling and allocation decisions. Simulation results demonstrate that our proposed algorithms significantly improve circuit execution time and scheduling efficiency (measured by makespan and throughput), while also reducing inter-QPU communication overhead, compared to baseline strategies. This work provides valuable insights into resource management strategies for scalable and heterogeneous DQC systems.
\end{abstract}

\begin{IEEEkeywords}
Distributed quantum computing, Quantum networks, Resource allocation, Quantum circuit scheduling
\end{IEEEkeywords}

%

\section{Introduction}
%
%
%
%
\IEEEPARstart{Q}{uantum} computing has gained attention as a solution for tackling intractable problems, due to its capacity to solve them significantly faster than traditional computers. In recent years, there have been notable advancements in quantum hardware and control systems, leading to the development of noisy intermediate-scale quantum processing units (QPUs). However, despite these efforts, current quantum processors still remain limited in their computational power \cite{cuomo2020towards,caleffi2024distributed}. Distributed quantum computing (DQC) aims to harness the collective power of multiple interconnected quantum processors, enabling the execution of larger and more complex quantum algorithms. In DQC, quantum algorithms are partitioned and executed across a network of quantum processors, which are interconnected through both quantum and classical communication channels. By distributing the computational workload among multiple quantum processors, this approach facilitates the development of scalable quantum computing systems that can surpass the limitations imposed by individual quantum processors \cite{yimsiriwattana2004distributed,barral2025review}.

\begin{figure}[t]
	\centering
        \includegraphics[width=0.95\linewidth]
        {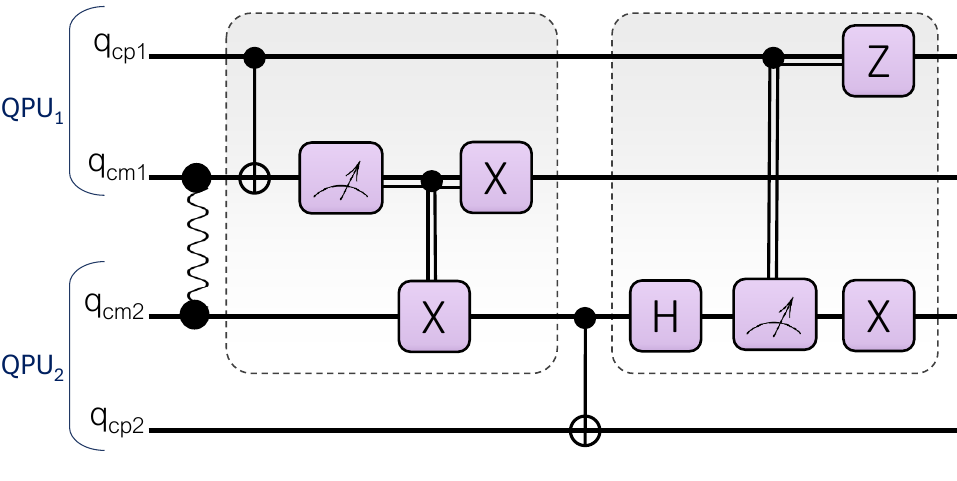}
        \vspace{-3mm}
	\caption{Circuit diagram of remote CNOT gate performed between two QPUs.} 
	\label{fig:remotegate}  
          \vspace{-0.4cm}
\end{figure}

DQC is expected to progress through stages of increasing scale and heterogeneity \cite{caleffi2024distributed}. This progression spans from the integration of multiple quantum processors within a single large quantum computer to the establishment of interconnected quantum processors across various quantum data centres. A quantum data centre, in this context, refers to a facility housing quantum computers at the scale of 10s - 100s meters, like a classical data centre. {One of the fundamental requirements for DQC is the ability to perform quantum operations between distant qubits located on separate QPUs, often referred to as \textit{remote gates}. Such operations typically rely on three components: remote entanglement generation, local quantum operations at each QPU, and classical communication between them \cite{yimsiriwattana2004distributed,campbell2024quantum}. As an illustrative example, Fig.~\ref{fig:remotegate} shows the circuit for implementing a controlled-NOT (CNOT) gate between two computing qubits, $q_{\rm cp 1}$ and $q_{\rm cp 2}$, residing on ${\rm QPU}_1$ and ${\rm QPU}_2$, respectively. The procedure begins with generating entanglement between dedicated communication qubits, $q_{\rm cm 1}$ and $q_{\rm cm 2}$, forming a shared entangled qubit pair, commonly referred to as an entangled bit (\textit{ebit}). The Pauli-X gate operation on ${\rm QPU}_2$ and the Z gate operation on ${\rm QPU}_1$ are then conditioned on classical measurement outcomes exchanged between the QPUs. This approach generalises to other controlled-unitary operations.

The integration of quantum networking and computation in DQC requires coordinated resource management across both domains. Compute resources refer to the computing qubits within QPUs, while communication resources include communication qubits and networking components such as optical switches. In this work, we assume that the network reliably establishes entanglement links to support remote operations and allocates communication resources on demand. Focusing on compute resource management, we conceptually distinguish two key components. The first jointly determines QPU allocation and the partition count per circuit under multiple concurrent circuits; we refer to this decision process as \textit{circuit scheduling}. The second component, \textit{circuit partitioning}, performs QPU--circuit mapping at a finer granularity by assigning the circuit qubits to the allocated QPUs according to the chosen partition count, while accounting for QPU capacities and the circuit’s interaction structure. In this work, we focus primarily on the scheduling layer, while explicitly capturing its interdependence with circuit partitioning by incorporating partition granularity (i.e., the number of partitions per circuit) as a decision variable in the optimisation framework.
}

{
Job scheduling in classical cloud data centres is a well-studied problem, typically focusing on objectives such as maximising resource utilisation, minimising makespan (i.e., the total completion time of all tasks), and balancing workload across computing resources \cite{hussain2018ralba,singhal2021resource,mishra2020load,elzeki2012overview}. In such environments, scheduling decisions primarily influence job execution time and system-level efficiency metrics. However, DQC introduces additional critical considerations. In particular, remote gate operations introduce infidelities that can degrade computational accuracy. These infidelities arise from qubit decoherence during remote gate execution as well as from imperfect entanglement generation between QPUs. Since scheduling decisions determine how circuits are distributed across QPUs, they directly influence the frequency and quality of inter-QPU entanglement link usage. Consequently, in DQC, scheduling decisions affect not only circuit execution time but also the computational fidelity of quantum tasks, inter-QPU communication overhead, and overall system-level performance metrics such as makespan. In addition, DQC scheduling must satisfy \emph{capacity constraints}, requiring that the selected QPUs provide sufficient combined physical qubits to meet a circuit’s qubit demand. This further increases the complexity of the decision process. Furthermore, prior studies \cite{oliveira2023fpga} have shown that quantum circuits exhibit varying degrees of sensitivity to errors introduced by distributed execution. For example, if a sparsely connected circuit is scheduled first and occupies the most favourable QPUs, such as those with high-fidelity links or larger qubit capacities, subsequent densely connected circuits may be forced into fragmented or suboptimal allocations. This can increase overall inter-QPU communication overhead and degrade execution performance. These characteristics motivate the design of circuit scheduling strategies that jointly consider qubit-capacity feasibility, circuit structure, and inter-QPU link fidelity under multi-circuit contention.

Beyond the considerations mentioned above, unlike classical scheduling where job execution time can typically be estimated prior to resource allocation, in DQC, the execution time of a quantum circuit depends on how it is partitioned across QPUs, which is itself determined by the scheduling decision. This tight coupling makes proper prior estimation challenging and motivates the adoption of lightweight surrogate models, the regression-based partitioning cost estimation in Section \ref{sec:QCirc_scheduling}.

While extensive research has addressed various aspects of DQC, including network resource management, quantum compilation, and circuit partitioning, comparatively fewer works have focused on QPU allocation and circuit scheduling. Several studies have addressed network resource management and scheduling, focusing on network-level operations such as entanglement request scheduling, quantum routing, and efficient entanglement generation or allocation protocols \cite{vista2025entanglement,zhang2025switchqnet,networkaware2025,cicconetti2022resource}. In contrast, our work addresses compute resource management, specifically Quantum
Circuit (QCirc) scheduling and QPU--QCirc mapping in heterogeneous DQC networks.

Quantum compilation and circuit partitioning for DQC have been studied extensively. Prior works have addressed compiler design challenges and optimisation of circuit depth and remote entanglement overhead under network and hardware constraints \cite{ferrari2021compiler,cuomo2023optimized,ferrari2023modular}. In parallel, graph- and hypergraph-based techniques have been proposed to partition QCircs so as to minimise remote entanglement overhead across distributed QPUs \cite{daei2020optimized,cambiucci2023hypergraphic,andres2019automated,davarzani2020dynamic,kaur2025optimized,dadkhah2021new}. Beyond partitioning, Andres et al.~\cite{andres2024distributing} considered circuit placement and distribution over heterogeneous networks with arbitrary topologies, integrating mapping decisions within a compiler framework. However, these approaches focus on optimising DQC for a single QCirc and do not explicitly address multi-circuit resource contention in heterogeneous DQC systems. Parekh et al.~\cite{parekh2021quantum} proposed a greedy scheduling algorithm that assigns QCircs sequentially to available QPUs. However, their approach does not jointly account for network topology, QPU decoherence properties, and circuit-specific structural characteristics within a unified optimisation framework.

Here, we address the problem of \textit{circuit scheduling} in DQC networks, accounting for quantum-specific challenges including fidelity degradation due to remote operations. We consider a general heterogeneous network model with QPUs of varying capacities, arbitrary connectivity topologies, and diverse QCircs. Within this framework, we propose two scheduling algorithms: (i) a dynamic batch scheduling approach that leverages mixed-integer linear programming (MILP) to jointly optimise QPU allocation and partition count for a batch of QCircs, and (ii) an MILP-based online scheduling strategy that allocates circuits individually. The proposed methods are analysed and evaluated through detailed simulations. Our results indicate that quantum-aware, batch-level optimisation can reduce inter-QPU communication overhead and improve system-level efficiency, particularly in heterogeneous network settings. The analysis also highlights a trade-off between makespan and communication cost, offering useful design insights for emerging quantum data centre architectures.
}

The remainder of this paper is organised as follows. Sec.~\ref{sec:DQC_network_model} describes the DQC network model. Section~\ref{sec:QCirc_scheduling} presents the proposed Batch-QCirc and Single-QCirc scheduling methods. Section~\ref{sec:evaluation} evaluates the proposed methods through simulations. Finally, Section~\ref{sec:conclusion} concludes the paper.

\section{DQC Network Model}
\label{sec:DQC_network_model}

\begin{figure}[t]
    \centering
  \includegraphics*[width=0.36\textwidth]{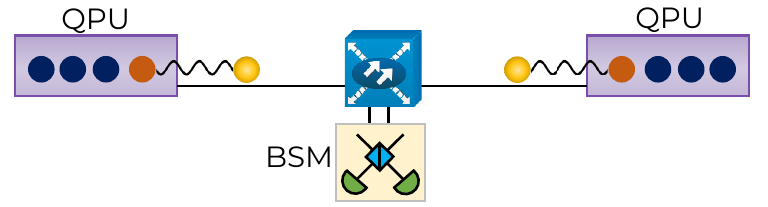}
  \vspace{-3mm}
    \caption{\small{High-level schematic of remote entanglement generation between a pair of QPUs via an intermediate BSM module. Blue circles denote computing qubits, and the brown circle denotes the communication qubit. }}
    \label{fig:REG}
    \vspace{-3mm}
\end{figure}

\begin{figure}[t]
    \centering
    \subfloat[]{\includegraphics*[width=0.5\textwidth]{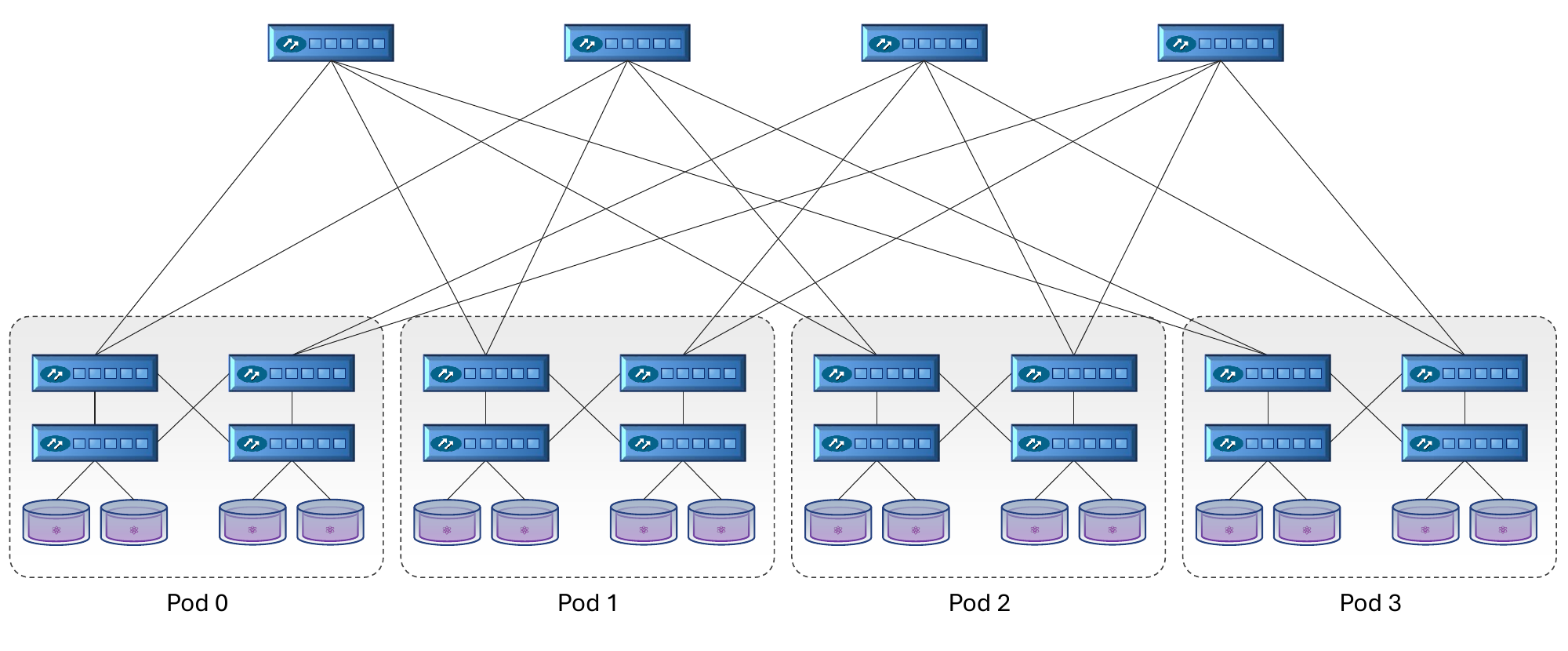}}\\
    
    \subfloat[]{\includegraphics*[width=0.45\textwidth,trim=0 60pt 0 0 pt,clip]{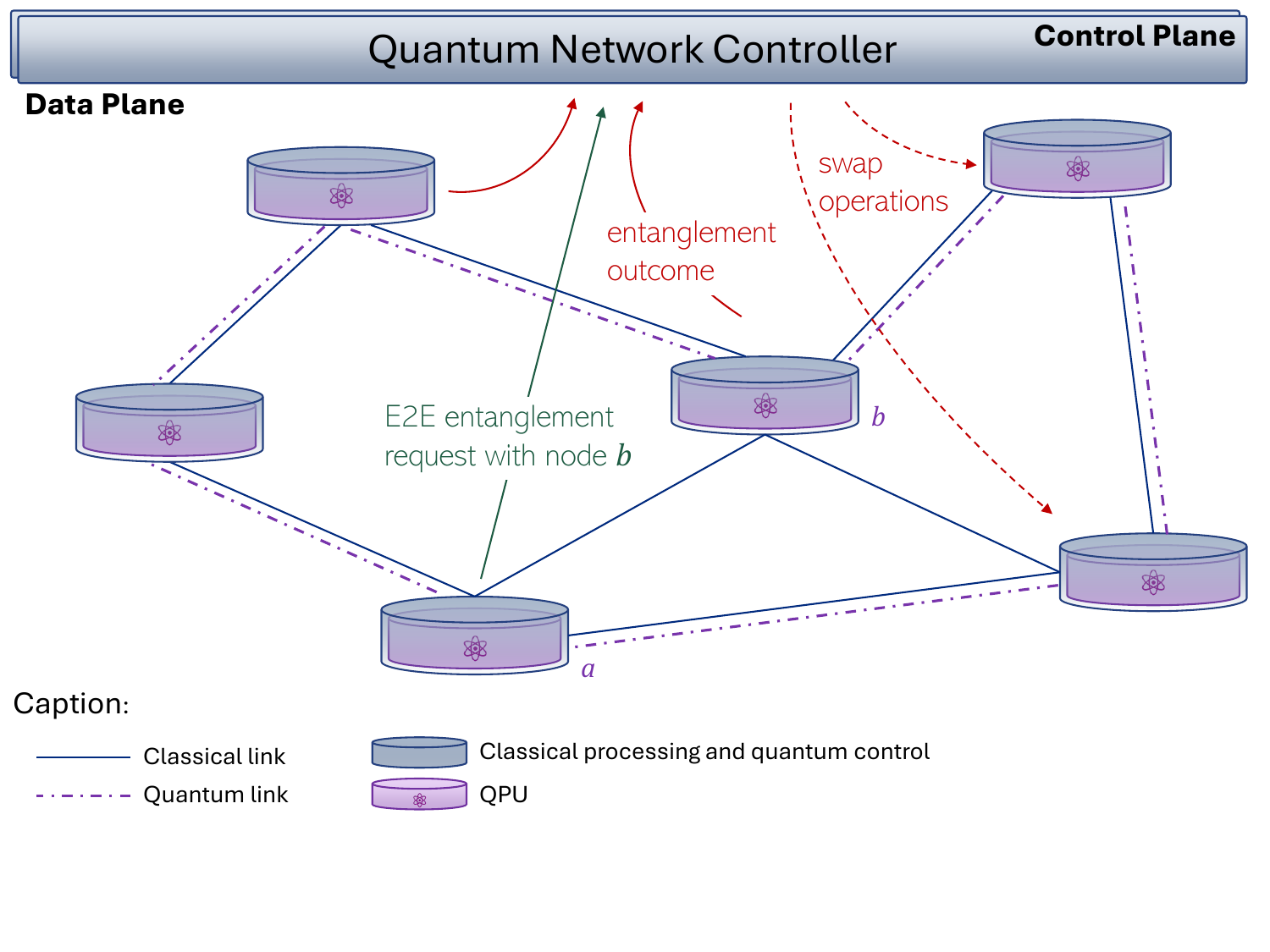}}
    \caption{\small{(a) Example network architecture for a DQC network within a quantum data centre (fat-tree topology with 4 pods). (b) High-level abstract representation of the DQC network model.}}
    \label{fig:quantumNetwork}
    \vspace{-3mm}
\end{figure}

We investigate a DQC network within a quantum data centre, where QPUs are interconnected via both quantum and classical links. Promising architectures for such data centre networks will employ low-loss high-speed optical switches to dynamically establish quantum links between QPUs. Quantum information of stationary computing qubits can be encoded in photons via transducers, using degrees of freedom such as time-bin, polarization, frequency, etc \cite{shapourian2025quantum, ang2024arquin}. At the physical and link layers, remote entanglement between two QPUs can be generated in various ways \cite{beukers2024remote}, with one example illustrated in Fig.~\ref{fig:REG}. This method relies on a Bell state measurement (BSM) module placed between the QPUs, enabling heralded entanglement generation. From an architectural perspective, switch-based data centre topologies such as fat-trees are promising candidates. A representative example with four pods is shown in Fig.~\ref{fig:quantumNetwork}(a), where QPUs are interconnected through three layers of optical switches to provide all-to-all connectivity. Alternatively, some designs may leverage entanglement swapping, allowing a QPU to act as a repeater and connect QPUs that are not directly linked.

At a higher abstracted connectivity level, the DQC network can be represented as in Fig.~\ref{fig:quantumNetwork}(b), which highlights the key components: QPUs connected by quantum and classical links, coordinated by a Quantum Network Controller. This abstraction captures the essential elements of the network while remaining agnostic to specific physical implementations and architectural details. 

\begin{figure*}[htbp]
   \centering   
    \subfloat{{\includegraphics[width=0.8\linewidth]{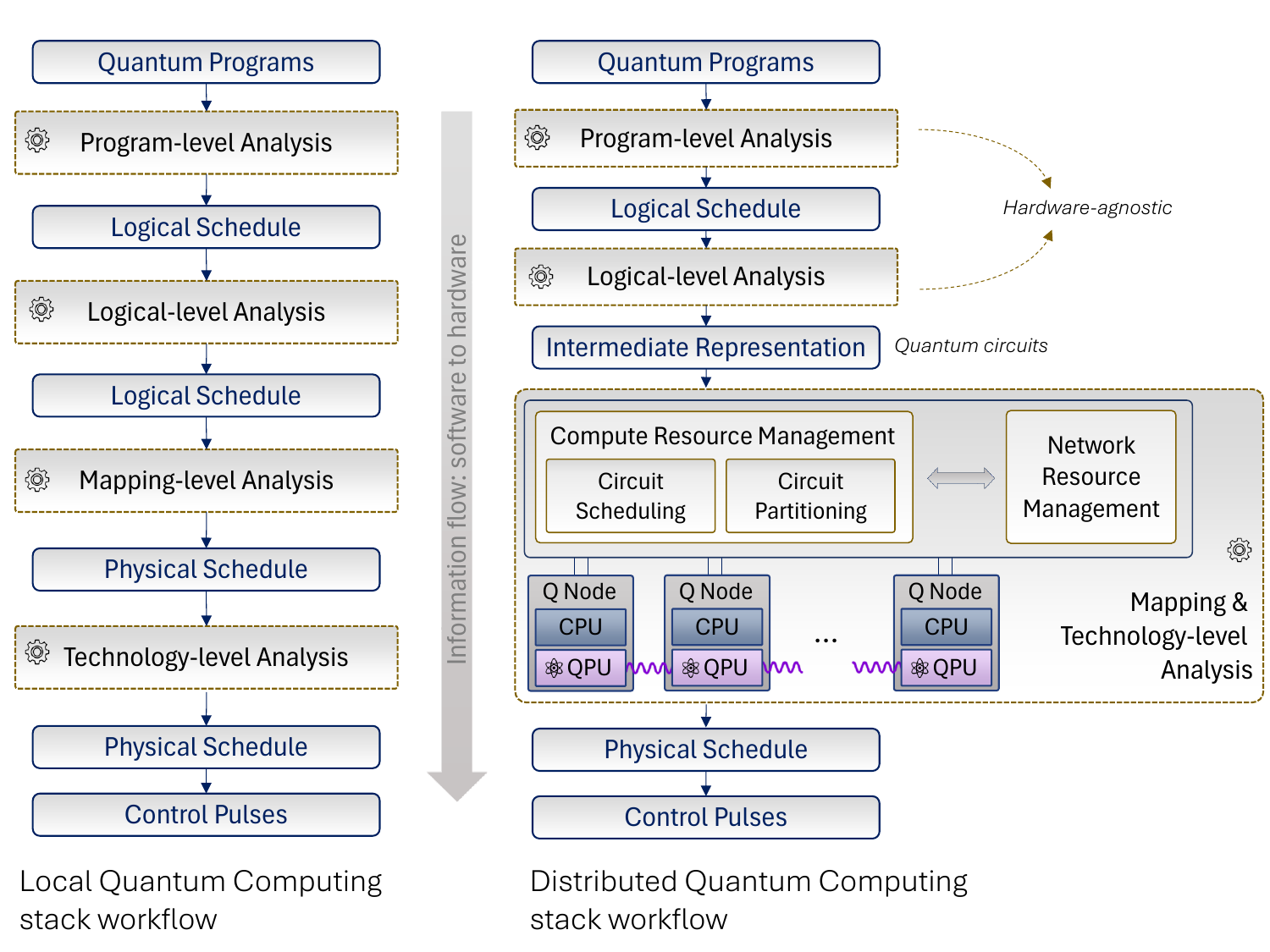}}}        
    \vspace{-3mm}   
    \caption{Workflow of the quantum stack for both local and distributed quantum computing. A layer-oriented approach for compilation tools that bridge quantum algorithms with quantum devices. The stack workflow for local quantum computing, depicted on the left side of the figure, is based on \cite{shi2020resource}. }
    \label{fig:computingstack}
    \vspace{-3mm}
\end{figure*}

{
On top of the physical and logical connectivity layers, the DQC architecture can be understood through its compilation and execution workflow, as illustrated in Fig.~\ref{fig:computingstack}. A core element of quantum computing is quantum compilation, which translates a high-level quantum program into hardware-level instructions \cite{maronese2022quantum,shi2020resource}. This translation is performed through a layered compilation stack (left side of Fig.~\ref{fig:computingstack}). In the distributed setting, additional tasks arise at the level of the intermediate quantum representation, which we refer to as QCirc (right side of Fig.~\ref{fig:computingstack}). These tasks involve resource management across both compute and network domains. Compute resources correspond to the physical qubits embedded in QPUs, whereas communication resources include communication qubits and networking components such as optical switches.
}

We now turn to the key definitions and assumptions related to our DQC network model. First of all, we abstract the network into a logical graph model \(G(V,E)\), where QPUs are represented as nodes and quantum logical links as edges. {An edge between $QPU_{j_1}$ and $QPU_{j_2}$ is characterised by the entanglement establishment latency $T_{j_1 j_2}$ (i.e., the average time required for one successful entanglement generation event between QPUs) and the fidelity $F_{j_1 j_2}$ of the generated entangled pair. This model relies on several simplifying assumptions. First, we assume on-demand entanglement generation: remote entanglement is initiated and completed at the start of a remote operation, and the resulting entangled pair is used immediately, eliminating the need for long-term entangled-pair storage. Second, the latency and fidelity parameters are assumed to be fixed for each logical link. Under these modelling assumptions, $T_{j_1 j_2}$ captures the primary communication-induced timing overhead of distributed execution. It is primarily determined by the entanglement generation rate, which typically decreases with increasing link loss.} For example, in the fat-tree topology depicted in Fig.~\ref{fig:quantumNetwork}(a), link loss increases with the number of optical switches in the path, thereby increasing remote gate execution time. Link's fidelity \(F_{j_1j_2}\) is similarly influenced by factors such as link loss and the specific entanglement generation protocol used.

The set of QPUs in the network is denoted by $\mathcal{QP} = \{ {\text{QPU}_j} \}_{j=1}^J$, where $J$ represents the total number of QPUs in the network. In the graph model \(G(V,E)\), the vertex set $V$ corresponds to $\mathcal{QP}$, i.e., $V \equiv \mathcal{QP}$. 
The capacity of ${\rm QPU}_j$, defined as its total number of computing qubits, is denoted by $N_j$. We also introduce a binary parameter $s_j$, which represents the current availability of ${\rm{QPU}}_j$. Specifically, $s_j=1$ if ${\rm{QPU}}_j$ is available, and $s_j=0$ otherwise. We assume that all QPUs are based on the same technology and exhibit nearly identical decoherence properties, characterised by a common decoherence time parameter, $T_{\text{dec}}$. It is important to note that the concept of ``decoherence time'' is technology-dependent. For example, in superconducting-qubit systems, two parameters, energy relaxation time $T_1$ and dephasing time $T_2$, are often reported; these can be combined into a single effective value, such as $\min(T_1, T_2)$ or $\frac{T_1 T_2}{T_1 + T_2}$, for modelling purposes. In trapped-ion systems, a memory lifetime parameter may be used instead. Regardless of the specific definition, there always exists a parameter (or suitable combination thereof) that captures the temporal limitations imposed by decoherence. To maintain platform independence, we abstract this into a single representative parameter $T_{\text{dec}}$ in our model.

\section{Proposed QCirc Scheduling methods}
\label{sec:QCirc_scheduling}
In this section, we consider the problem of QCirc scheduling for the DQC network described above. Our primary objective is to schedule a set of quantum circuits and allocate a subset of QPUs to each of them. The set of input quantum circuits is denoted by \( \mathcal{QC} = \{ \text{QCirc}_m \}_{m=1}^M \), where \( M \) is the total number of circuits. The required number of qubits for each circuit \( \mathcal{QC} = \{ \text{QCirc}_m \}\) is represented by \(w_m\). {We define $K_m^{\max}$ as the maximum allowable number of partitions for $\text{QCirc}_m$, determined according to factors such as data centre policy or circuit size. This parameter bounds circuit fragmentation, limits excessive inter-QPU communication, and captures practical fidelity constraints and system-level policy requirements.} Throughout this work, we assume a non-preemption model, meaning that once a QCirc begins execution on its allocated QPUs, it runs to completion without interruption. Furthermore, each QPU can host at most one partition at any given time, meaning it cannot concurrently execute partitions from different circuits.
In the following, we propose two primary QCirc scheduling methods. 
The first method, termed \textit{Batch-QCirc Scheduling}, is based on dynamic batch scheduling, where a group of circuits is selected and assigned to QPUs collectively. 
The second method, referred to as \textit{Single-QCirc Scheduling}, assigns each circuit individually. 

\begin{figure}[t]
	\centering
        \includegraphics[width=1.0\linewidth]{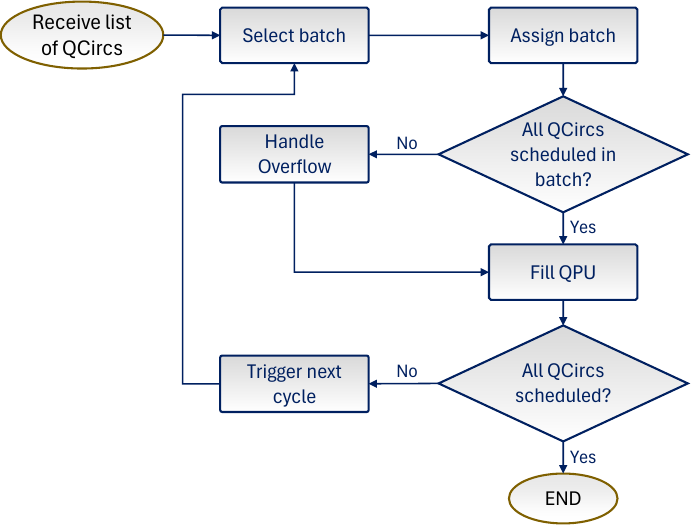}        
        \vspace{-5mm}
	\caption{Workflow of the Dynamic Batch-QCirc Scheduling method.} 
	\label{fig:batch_scheduling}  
\end{figure}
\vspace{-3mm}
\subsection{Dynamic Batch QCirc Scheduling}
\label{sec:batch_scheduler}
To efficiently schedule diverse quantum circuits (e.g. quantum Fourier transform), with varying qubit requirements and structures, we propose a batch scheduling framework that jointly optimises QPU allocation and the number of partitions for a batch of circuits. The algorithm operates in repeated scheduling cycles, each consisting of four main stages: (a) batch selection, where incoming quantum circuits are grouped into a batch based on their combined qubit requirements; (b) batch assignment, where the selected batch is mapped to available QPUs via a holistic MILP-based optimisation that accounts for decoherence effects and inter-QPU connectivity; (c) overflow handling, where any circuits that could not be assigned in the main step are promptly scheduled to run as soon as resources become available; and (d) filling idle QPUs, where any QPUs that remain unallocated after the batch assignment process are identified and, if suitable candidates exist in the remaining circuits of $\mathcal{QC}$, assigned to them, enhancing the utilisation of available resources.

The workflow for the proposed batch scheduling method is illustrated in Fig.~\ref{fig:batch_scheduling}, and the corresponding algorithm pseudocode is outlined in Algorithm~\ref{alg:batch_qcirc_scheduler}. 
In each scheduling cycle, a batch of circuits is selected from \( \mathcal{QC} \). 
This batch is denoted by \( \mathcal{B} = \{ \text{QCirc}_m \mid m \in \mathcal{I} \} \), where \( \mathcal{I} \subseteq \{1, \dots, M\} \) is the set of indices corresponding to the selected circuits. 
The batch size, \( S = |\mathcal{I}| \), may vary across scheduling cycles.

We adopt a simple batch selection process, where elements of \( \mathcal{QC} \) are sequentially added to \( \mathcal{B} \) based on their requests' arrival time until the total required number of qubits, \( c_{\rm req}=\sum_{m \in \mathcal{I}} {w_m}\), exceeds a threshold \( \beta c_{\rm tot} \). 
Here, \( c_{\rm tot} \) denotes the total qubit capacity of all available QPUs, given by \( c_{\rm tot} = \sum_j {s_j N_j} \). 
The parameter \( \beta \), fixed in the range \( 0 < \beta < 1 \), is chosen to ensure that, with high probability, all circuits in the selected batch can be successfully assigned to available QPUs.

The selected batch then undergoes a batch assignment process, referred to as \textsc{AssignBatch} in Algorithm~\ref{alg:batch_qcirc_scheduler}, where the circuits are mapped to the available QPUs. 
We have developed an MILP-based batch assignment algorithm, which will be described in detail in Sec.~\ref{subsubsec:batch_assignment}.

After the batch assignment, there might be a few circuits unassigned if the \( \beta \) is large, e.g., close to 1, or some circuits are complex to be distributed among available QPUs given the constraints. Any circuits in the batch that remain unassigned are handled after batch assignment using the \textsc{HandleOverflow} procedure, described in Algorithm~\ref{alg:handle_overflow}. The QPUs are first sorted in ascending order based on their expected busy time (EBT). The unassigned circuits are then assigned sequentially to the sorted QPUs, while taking into account circuit constraints such as capacity and the maximum number of partitions. The EBT for a given QPU can be estimated from the execution time of circuits already assigned to it or, when such information is not available, from other factors such as circuit connectivity, the number of partitions assigned to the circuit, and historical assignment data. This overflow step acts primarily as a failsafe, as \(\beta\) is chosen to make such cases rare.

To improve QPU utilisation, we incorporate a filling mechanism that assigns circuits to idle, unallocated QPUs. The \textsc{FillQPU} algorithm, described in Algorithm~\ref{alg:fill_qpu}, selects circuits from \(\mathcal{QC}\) and assigns each of them to idle QPUs individually, using a MILP-based optimisation strategy. To reduce the risk of excessive remote gate generation, only circuits with a $\nu _m$ value below a specified threshold are considered. Here, \(\nu_m\) (see also Sec.~\ref{subsubsec:batch_assignment}) is a score parameter that quantifies the circuit's connectivity and its likelihood of generating costly remote gates in a distributed setting. The MILP-based Single-QCirc assignment process (\textsc{AssignQcirc} in Algorithm~\ref{alg:fill_qpu}) is detailed in Sec.~\ref{sec:single_scheduler}.

Once all QPUs are allocated, the algorithm waits until a fraction \(\alpha\) of the total qubit capacity across all QPUs, i.e., \(\alpha\sum_j{N_j}\), becomes available, then proceeds to the next scheduling cycle. This mechanism is referred to as \textsc{TriggerNextCycle} in Algorithm~\ref{alg:batch_qcirc_scheduler}.

{The parameters $\beta$ and $\alpha$ control two distinct phases of the 
scheduling cycle with complementary roles in shaping system-level performance. 
The parameter $\beta$ controls batch formation prior to optimisation by 
limiting the aggregate qubit demand relative to the available system capacity, 
thereby reducing the likelihood of infeasibility, while $\alpha$ determines 
the temporal progression of successive scheduling cycles. Both parameters 
affect the scope of joint optimisation within a cycle. Setting these parameters 
appropriately is therefore important: too large a $\beta$ risks infeasibility, 
while too large an $\alpha$ introduces unnecessary idle time between cycles, 
as confirmed by the simulation results in Sec.~\ref{sec:evaluation}, where 
$\alpha = 0.75$ yields worse makespan than $\alpha = 0.55$.

}

\begin{algorithm}[t]
\caption{Dynamic Batch-QCirc Scheduler}
\label{alg:batch_qcirc_scheduler}
\begin{algorithmic}[1]
    \State \textbf{Input:} $\mathcal{QP}$, $\mathcal{QC}$, $G$
    \While{not \textsc{IsEmpty}($\mathcal{QC}$)}
        \State \textit{batch} $\gets$ \textsc{SelectBatch}($\mathcal{QC},\beta$)
        \State \textit{batchMap} $\gets$ \textsc{AssignBatch}(\textit{batch}, $\mathcal{QP}$,$G$)
        \If{not \textsc{AllAssigned}(\textit{batch}, \textit{batchMap})}
            \State \textsc{HandleOverflow}(\textit{batch},\textit{batchmap}, $\mathcal{QP}$)
        \EndIf
        \State \textsc{RemoveAssigned}($\mathcal{QC}$, \textit{batch})
        \If{\textsc{AnyAvailable}($\mathcal{QP}$)}
            \State \textsc{FillQPU}($\mathcal{QC}$, $\mathcal{QP}$)
        \EndIf
        \State \textsc{TriggerNextCycle}($\mathcal{QP},\alpha$)
    \EndWhile
\end{algorithmic}
\end{algorithm}

\begin{algorithm}[htbp]
\caption{\textsc{HandleOverflow}}
\label{alg:handle_overflow}
\begin{algorithmic}[1]
    \State \textbf{Input:} \textit{batch}, \textit{batchMap}, $\mathcal{QP}$, $G$
    \State $U \gets$ Unassigned QCircs in \textit{batch} 
    \State $\mathcal{QP}_{\rm s} \gets$ QPUs in $\mathcal{QP}$ sorted by increasing EBT
    \For{$\text{QCirc}_m$ in $U$}
        \State Scan $\mathcal{QP}_{\rm s}$ to find a set 
        $\{\text{QPU}_{j_k}\}_{k=1}^K$
        \Statex \quad such that $\sum_{k=1}^K N_{j_k} \geq w_m$ and $K \leq K^{\max}_m$
        \State Assign $\text{QCirc}_m$ to $\{\text{QPU}_{j_k}\}_{k=1}^K$
        \State $\mathcal{QP}_{\rm s} \gets \mathcal{QP}_{\rm s} \setminus \{\text{QPU}_{j_k}\}_{k=1}^K$\Comment{remove allocated QPUs}
    \EndFor
\end{algorithmic}
\end{algorithm}

\begin{algorithm}[htbp]
\caption{\textsc{FillQPU}}
\label{alg:fill_qpu}
\begin{algorithmic}[1]
    \State \textbf{Input:} $\mathcal{QC}$, $\mathcal{QP}$, $G$
    \State \textit{avlCapacity} $\gets \sum_j N_j$ for idle QPUs
    \If{\textit{avlCapacity}$\;>0$}
        \For{each $\text{QCirc}_m$ in $\mathcal{QC}$}
            \If{$\nu_m \leq \Gamma$ \textbf{and} $w_m \leq$ \textit{avlCapacity}}
                \State \textsc{AssignQCirc}($\text{QCirc}_m$)
                \State \textsc{RemoveAssigned}($\text{QCirc}_m$)
                \State Update \textit{avlCapacity}
            \EndIf
        \EndFor
    \EndIf
\end{algorithmic}
\end{algorithm}

\subsubsection{Batch-QCirc Assignment}
\label{subsubsec:batch_assignment}
We propose an optimisation algorithm to minimise the detrimental effects of inter-QPU communication, focusing on decoherence-induced errors and fidelity loss from remote entanglement generation. Remote gates incur significantly higher execution time than local gates due to quantum and classical communication delays and loss, increasing qubit exposure to decoherence. Moreover, remote entanglement generation is inherently imperfect, with fidelity that can degrade further through operations such as entanglement swapping and distillation. Repeated use of remote gates amplifies these effects, causing substantial fidelity loss and degrading the overall computational accuracy of distributed quantum computations. Our approach mitigates this degradation for the batch of QCircs as a group. The output of the Batch-QCirc assignment specifies the number of partitions assigned to each QCirc and the QPUs allocated for their execution, which then guide the circuit partitioning process in efficiently decomposing each QCirc into smaller sub-circuits.
To mathematically formulate the problem of Batch-QCirc assignment, we begin by defining the following decision variables:
\begin{itemize}
\item Let $r_{mj}$ be a binary variable, where $r_{mj}=1$ if ${\rm QPU}_j$ is allocated to ${\rm QCirc}_m$, and $r_{mj}=0$ otherwise.
\item Let $y_{mk}$ be a binary variable, where $y_{mk}=1$ if ${\rm QCirc}_m$ is partitioned into exactly $k$ parts (i.e., allocated to $k$ distinct QPUs), and $y_{mk}=0$ otherwise.
\end{itemize}
We define a cost function that captures the impact of qubit decoherence and imperfect entanglement generation during remote gate executions as follows:
\begin{align}
C &= \sum_{m \in \mathcal{I}} \sum_{k=1}^{K_m^{\rm max}} 
      \sum_{1 \leq j_1 < j_2 \leq J} 
      \nu_{mk}\;y_{mk}\;r_{mj_1}r_{mj_2} \big(\omega_0 \frac{w_m T_{j_1 j_2}}{T_{\rm dec}}
      + \nonumber \\
    &\qquad\omega_1 (1 - F_{j_1 j_2})\big) 
\label{cost_function}
\end{align}
where $T_{j_1 j_2}$ is the latency parameter, i.e., average time required to establish entanglement, associated with a remote gate between ${\rm QPU}_{j_1}$ and ${\rm QPU}_{j_2}$, and $F_{j_1 j_2}$ is the fidelity parameter associated with remote entanglement generation for the QPU pair. The term $(w_m T_{j_1 j_2}) / T_{\rm dec}$ captures the relative decoherence impact, where larger circuits (\(w_m\)) and longer communication delays (\(T_{j_1 j_2}\)) increase decoherence relative to the characteristic decoherence parameter \(T_{\rm dec}\). The parameters $\omega_0$ and $\omega_1$ are constant weights that reflect the relative importance of decoherence‑induced errors versus fidelity loss from entanglement generation; unless otherwise stated, we set $\omega_0 = \omega_1$.

\begin{table}
\caption{\small{List of key notations and parameters}} 
\vspace{-2mm}
\centering
    \begin{tabularx}{\linewidth}{cX}
    \toprule
        \textbf{Parameter} & \textbf{Definition}\\
      \midrule
$G(V,E)$& Graph representation of DQC network\\
$\mathcal{QP}$ & Set of all QPUs\\
$\mathcal{QC}$ & Set of all QCircs\\
$\mathcal{B}$ & Selected batch of QCircs\\
$J$ & Number of QPUs\\
$M$ & Number of QCircs\\
$S$ & Number of QCircs in the selected batch\\
$T_{\text{dec}}$ & Decoherence time parameter of QPUs\\
$N_j$ & Number of computing qubits within ${\rm{QPU}}_j$\\
$s_j$ & Equal to 1 if ${\rm{QPU}}_j$ is available, and 0 otherwise\\ 
$w_m$& Number of qubits required for ${\rm{QCirc}}_m$\\
$K_m^{\rm max}$& Partition count threshold for ${\text{QCirc}}_m$\\
$r_{mj}$& Binary variable equal to 1 if ${\text{QCirc}}_m$ is assigned to ${\text{QPU}}_j$, and 0 otherwise\\
$y_{mk}$ & Binary variable equal to 1 if ${\text{QCirc}}_m$ is partitioned to $k$ parts, and 0 otherwise\\
$T_{j_1 j_2}$& The latency parameter for establishing entanglement for remote gate execution, involving ${\rm{QPU}}_{j_1}$ and ${\rm{QPU}}_{j_2}$\\
$F_{j_1 j_2}$& The fidelity parameter associated with remote entanglement generation between ${\rm{QPU}}_{j_1}$ and ${\rm{QPU}}_{j_2}$\\
$\nu_{mk}$ & Partitioning cost coefficient for ${\text{QCirc}}_m$ and a given partition count $k$\\ 
$g _m$ & Average degree of nodes in the graph model of ${\text{QCirc}}_m$\\
$\sigma_m$ & Std. dev. of node degrees in the graph model of ${\text{QCirc}}_m$\\
$\lambda _{2,m}$& Algebraic connectivity of the graph model of ${\text{QCirc}}_m$\\
$\gamma _{m}$& Weighted density of the graph model of ${\text{QCirc}}_m$\\
      \bottomrule
    \end{tabularx}
    \label{parameters}
    \vspace{-2mm}
\end{table}

To incorporate circuit structure and connectivity patterns into our MILP objective, we introduce a partitioning cost coefficient $\nu_{mk}$, {which captures the structural overhead induced by distributing \(\text{QCirc}_m\) across $k$ QPUs} and thereby enables circuit-structure-aware resource allocation. Each quantum circuit is represented as a weighted undirected graph, where nodes correspond to qubits and edges correspond to two-qubit gates. The weight of each edge is given by the number of occurrences of gates between the respective qubit pair. Based on this representation, $\nu_{mk}$ is defined as the estimated graph cut size under $k$-way partitioning (with partitions of equal or nearly equal size). Although many distinct $k$-partitions are possible, adopting a standard balanced $k$-partition is sufficient to capture a circuit’s connectivity level and its tendency to generate costly remote gates, while avoiding extra complexity.

To efficiently obtain $\nu_{mk}$ for different circuits without performing explicit graph partitioning, we propose an estimation method based on a linear regression model trained on three structural graph features: weighted density, algebraic connectivity, and coefficient of variance. This regression provides a lightweight estimate of cut size across circuit types, thereby making $\nu_{mk}$ tractable to compute in our MILP formulation. The three features are defined as follows:  

(i) \textbf{Weighted density} ($\gamma_m$) quantifies the average connectivity per qubit pair, computed as  
\begin{align}
\gamma_m = \frac{\tfrac{1}{2}\sum_{i=1}^{w_m} g_m^{(i)}}{\binom{w_m}{2}},
\end{align} 
where $g_m^{(i)}$ denotes the weighted degree of qubit node $i$. The numerator corresponds to the total edge weight of the circuit graph, i.e., the total number of two-qubit gate occurrences across the circuit. 

(ii) \textbf{Algebraic connectivity} ($\lambda_{2,m}$) is the second-smallest eigenvalue of the normalised Laplacian matrix $L_m$ of the circuit graph \cite{fiedler1973}. It measures global connectivity strength; larger values indicate that the graph is harder to partition without cutting many edges.  

(iii) \textbf{Coefficient of variance} ($\sigma_m/g_m$) captures connectivity imbalance across qubits, defined as  
\begin{align}
\frac{\sigma_m}{g_m} = \frac{\sqrt{\frac{1}{w_m}\sum_{i=1}^{w_m}(g_m^{(i)} - g_m)^2}}{\frac{1}{w_m}\sum_{i=1}^{w_m} g_m^{(i)}},
\end{align}  
where $g_m$ and $\sigma_m$ are the average and standard deviation of the weighted degrees $\{g_m^{(i)}\}$, respectively.  
 
\noindent These features jointly capture complementary aspects of circuit structure: 
$\gamma_m$ reflects the overall connectivity strength, particularly the density of pairwise interactions; 
$\lambda_{2,m}$ indicates how robustly the graph remains connected under partitioning; 
and $\sigma_m/g_m$ captures connectivity imbalance across qubits that may create partitioning bottlenecks. 
Importantly, these features scale smoothly with circuit size $w_m$, enabling the simple but effective \textbf{linear regression model} to learn structural tendencies rather than raw size.

The regression target is the normalised $k$-way cut size:  
\begin{equation}
\tilde{\nu}_{mk} = \frac{{\rm Cut}_m^{(k)}}{\frac{1}{2}\sum_{i=1}^{w_m} g_m^{(i)}},
\label{cutsize}
\end{equation}
where ${\rm Cut}_m^{(k)}$ is the sum of edge weights for inter-partition edges when circuit $m$ is optimally partitioned into $k$ balanced parts. We then rescale by the total edge weight to recover the estimated partitioning cost coefficient:  
\begin{equation}
\nu_{mk} \approx \Big({\hat{\chi}}_{0,k}\gamma_m+ {\hat{\chi}}_{1,k}\lambda_{2,m} + {\hat{\chi}}_{2,k}\tfrac{\sigma_m}{g_m} + {\hat{\chi}}_{3,k} \Big) \cdot \big({\tfrac{1}{2}\sum_{i=1}^{w_m} g_m^{(i)}}\big).
\label{LR}
\end{equation}  
Here, $({\hat{\chi}}_{0,k},{\hat{\chi}}_{1,k},{\hat{\chi}}_{2,k},{\hat{\chi}}_{3,k})$ denote the regression coefficients, which are pre-learned and fitted by minimising the mean squared error (MSE)  
\begin{equation}
\frac{1}{|\mathcal{D}|} \sum_{d\in \mathcal{D}}
\left(
\tilde{\nu}_{dk} - 
\big(\hat{\chi}_{0,k} \gamma_d+ \hat{\chi}_{1,k} \lambda_{2,d}+ \hat{\chi}_{2,k}\tfrac{\sigma_d}{g_d} + \hat{\chi}_{3,k}\big)
\right)^2,
\end{equation}
for a given $k$ partitions. In this expression, $\mathcal{D}$ denotes a set of quantum circuits with varying types and qubit counts, and $\tilde{\nu}_{dk}$ is the normalised $k$-way cut size for circuit $d \in \mathcal{D}$.

Our optimisation problem, which minimises the cost function $C$ in (\ref{cost_function}), is formulated as the MILP in Formulation~\ref{milp_formulation}.  
The cost function is linearised by introducing the auxiliary variables $z_{mj_1 j_2} = r_{mj_1} r_{mj_2}$ and $x_{m k j_1 j_2} = z_{mj_1 j_2} y_{m k}$.  
The role of each constraint in Formulation~\ref{milp_formulation} is as follows:

\noindent{(1) Minimum assignment:}  
Enforces a lower bound on the number of circuits scheduled in the current cycle.  
By default, $\zeta = S$ so that all circuits in the batch are assigned.  
If the MILP becomes infeasible, $\zeta$ is decremented until a feasible allocation is found.

\noindent{(2) Capacity requirement:}  
Ensures that the aggregate qubit capacity of the QPUs assigned to ${\rm QCirc}_m$ meets or exceeds its qubit requirement $w_m$. While $s_j$ (QPU availability binary parameter) may vary between scheduling cycles, it is fixed within each cycle and thus treated as a parameter in the MILP.  

\noindent{(3) Single‑partition per QPU:}  
Ensures that each QPU is assigned at most one circuit partition in a scheduling cycle.

\noindent{(4) Link between binary variables:}  
Provides an MILP‑compatible link between $r_{mj}$ and $y_{mk}$.  

\noindent{(5)–(6) Linearisation of auxiliary variables:}  
Implements MILP‑compatible linear formulations of the equalities $z_{mj_1 j_2} = r_{mj_1} r_{mj_2}$ and $x_{m k j_1 j_2} = z_{mj_1 j_2} y_{mk}$.

\begin{formulationN}{MILP-based Batch-QCirc assignment problem}
\begin{equation*}
\begin{aligned}
&\min~ \bigg\{ 
      \sum_{m \in \mathcal{I}} \sum_{k=1}^{K_m^{\rm max}} 
      \sum_{1 \leq j_1 < j_2 \leq J} 
      \nu_{mk}\, x_{mkj_1 j_2} \big(\omega_0 \frac{w_m T_{j_1 j_2}}{T_{\rm dec}}
      + \\
    &\qquad\qquad 
       \omega_1 (1 - F_{j_1 j_2})
      \big) 
    \bigg\} \\
    &\text{s.t.} \\
&(1)~\sum_{m \in \mathcal{I}} \sum_{k=1}^{K_m^{\rm max}} y_{mk}= \zeta \\
&(2)~\sum_{j=1}^{J} r_{mj} s_j N_j \;\geq\; w_m \sum_{k=1}^{K_m^{\rm max}} y_{mk} \quad \forall m \in \mathcal{I}\\
&(3)~\sum_{m \in \mathcal{I}} r_{mj} \;\leq\; 1 \quad \forall j\\
&(4)~\sum_{j=1}^{J} r_{mj} \;=\; \sum_{k=1}^{K_m^{\rm max}} k\,y_{mk} \quad \forall m \in \mathcal{I}\\
&(5)~ r_{m j_1}+r_{m j_2}-1 \;\leq\; z_{m j_1 j_2} \;\leq\; r_{m j_1},\; r_{m j_2}\\
&\qquad \forall m \in \mathcal{I},~ j_1<j_2 \\
&(6)~ z_{m j_1 j_2}+y_{mk}-1 \;\leq\; x_{m k j_1 j_2} \;\leq\; z_{m j_1 j_2},\; y_{mk}\\
&\qquad \forall m \in \mathcal{I},~ k,~ j_1<j_2
\end{aligned}
\end{equation*}
\label{milp_formulation}
\end{formulationN}

\begin{algorithm}[b]
\caption{Single-QCirc Scheduler}
\label{alg:single_qcirc_scheduler}
\begin{algorithmic}[1]
    \State \textbf{Input:} $\mathcal{QC}$, $\mathcal{QP}$, $G$ 
    \While{not \textsc{IsEmpty}($\mathcal{QC}$)}
        \For{each $\text{QCirc}_m$ in $\mathcal{QC}$}
            \State \textsc{AssignQCirc}($\text{QCirc}_m$)
            \State \textsc{RemoveAssigned}($\text{QCirc}_m$)
        \EndFor
    \EndWhile
\end{algorithmic}
\end{algorithm}

In the \textsc{FillQPU} algorithm (Algorithm~\ref{alg:fill_qpu}), we employ the parameter $\nu_{mk}$ as a scoring metric for circuit connectivity and remote-gate overhead estimation. In particular, we consider the case $k=2$ (with the index $k$ omitted in Algorithm~\ref{alg:fill_qpu} for simplicity), and only circuits with $\nu_{m}$ not exceeding a threshold $\Gamma$ are selected for individual assignment. {This prevents highly connected circuits from being handled by \textsc{FillQPU}, which lacks an optimisation mechanism, and reduces the risk of excessive remote gate overhead at this stage.}

{
As can be seen from the Batch-QCirc scheduling workflow, the framework incorporates a multi-layer feasibility handling mechanism to ensure robustness. Although batch selection limits the total qubit demand to a fraction $\beta$ of the combined capacity of the available QPUs, infeasibility may still arise due to MILP constraints such as partition count limits. To address this, the MILP includes the adaptive parameter $\zeta$, which specifies the number of circuits to be assigned in the current batch. If the problem is infeasible with $\zeta$ equal to the batch size, $\zeta$ is iteratively reduced until a feasible solution is obtained. Any remaining circuits are subsequently processed by the \textsc{HandleOverflow} procedure (Algorithm~2). Empirically, $\beta$ can be chosen to make such overflow events rare; in the experiments reported in Section~IV, $\beta=0.85$ was found to be a good operating point. Together, these steps guarantee that the scheduler consistently produces a feasible allocation and avoids deadlock.

It is worth noting that while this work assumes a homogeneous decoherence 
parameter $T_{\rm dec}$ across all QPUs, the framework can be naturally 
extended to systems where QPUs exhibit heterogeneous decoherence times. In particular, $T_{\rm dec}$ in the objective function in~\eqref{cost_function} can be replaced with an effective coherence time, where for ${\rm QCirc}_m$ 
partitioned into $k$ parts:
\begin{equation}
\frac{1}{T^{{\rm eff}}_{mk}} = \frac{1}{k}\sum_{j} \frac{r_{mj}}{T^{{\rm dec}}_j},
\end{equation}
which is the harmonic mean of the decoherence times of the $k$ allocated 
QPUs. This captures the fact that a remote gate between a QPU pair 
introduces decoherence to all of the circuit's qubits on the allocated QPUs. 
Such a replacement preserves the linear structure of the MILP, since $1/k$ 
is a precomputed constant for each $k$, and the multiplicative binary 
variable $r_{mj}$ can be linearised following the same approach described 
above. A detailed quantitative evaluation of heterogeneous decoherence 
settings constitutes an interesting direction for future investigation.
}

\subsection{Single-QCirc Scheduling}
\label{sec:single_scheduler}
Another approach for scheduling quantum circuits is to schedule them individually rather than in batches. In this case, we propose a Single‑QCirc scheduling algorithm, presented in pseudocode in Algorithm~\ref{alg:single_qcirc_scheduler}. This algorithm uses an MILP‑based assignment method to optimise QPU allocation for each circuit $\rm{QCirc}_m$, taking into account the available QPUs, network topology, and link characteristics. The key difference from the Batch‑QCirc scheduling method described in the previous subsection is that the batch approach jointly optimises allocation for all circuits in a batch, whereas the single‑circuit method follows a one‑by‑one, online strategy that locally optimises each assignment. As a result, diversity in circuit structures and characteristics across the batch is not fully exploited in this method. On the other hand, the single‑QCirc approach offers lower computational complexity and enables immediate scheduling of each circuit as soon as the required resources become available. 

We adopt the same MILP‑based framework to determine the optimal allocation for each circuit. This assignment procedure is referred to as \textsc{AssignQCirc} in Algorithm~\ref{alg:single_qcirc_scheduler}, and its MILP formulation is provided in Formulation~\ref{milp_formulation_single}. The same objective function and constraints from the batch formulation are used, but simplified for the single‑circuit case. Note that \textsc{AssignQCirc}  is also used within the \textsc{FILLQPU} procedure described in Algorithm~\ref{alg:fill_qpu}.

\begin{formulationN}{MILP-based Single-QCirc assignment problem}
\begin{equation*}
\begin{aligned}
&\min~ \bigg\{ 
      \sum_{1 \le j_1 < j_2 \leq J} 
      z_{m j_1 j_2} \big(
        \omega_0 \frac{w_m T_{j_1 j_2}}{T_{\rm dec}}
        + \omega_1 (1 - F_{j_1 j_2})
      \big)
    \bigg\} \\
    &\text{s.t.} \\
& (1)~\sum_{j=1}^{J} r_{mj} s_j N_j \;\geq\; w_m, \quad (2)~\sum_{j=1}^{J} r_{mj} \;\leq\; K_m^{\rm max}\\
&(3)~ r_{m j_1}+r_{m j_2}-1 \;\leq\; z_{m j_1 j_2} \;\leq\; r_{m j_1},\; r_{m j_2},\quad \forall j_1<j_2 \\
\end{aligned}
\end{equation*}
\label{milp_formulation_single}
\end{formulationN}

\section{Evaluation and results}
\label{sec:evaluation}

This section evaluates the proposed scheduling algorithms through simulations. The scheduling methods (Algorithms~\ref{alg:batch_qcirc_scheduler} and~\ref{alg:single_qcirc_scheduler}) are implemented in Python. The MILP optimisation problems defined in Problem Formulations~1 and~2 are solved using the \texttt{Python-MIP} package with the Gurobi optimiser. The batch selection parameter $\beta$ in Algorithm~\ref{alg:batch_qcirc_scheduler} is set to $0.85$, and the maximum partition count threshold $K^{\text{max}}_m$ is fixed at $4$ for all circuits.

We first describe the evaluation setup, including the network model, benchmark circuits, workload scenarios, and scheduling–partitioning configurations. We then present and analyse the corresponding simulation results.

\subsection{Evaluation Setup}

\subsubsection{Network Model}

We consider a DQC network comprising 16 QPUs within a data centre. The QPUs are heterogeneous, with four QPUs of capacity 8 qubits, four of capacity 12 qubits, four of capacity 16 qubits, and four of capacity 20 qubits. These capacities are randomly assigned to individual QPUs using a fixed random seed to ensure consistent QPU configurations across all simulations. The entire simulation process is repeated for 10 different seed values, and the average results are presented.

{The network adopts a fat tree topology of four pods with each pod 
consisting of 4 QPUs, as illustrated in Fig.~\ref{fig:quantumNetwork}(a). 
Since fibre lengths within a data centre are short (ranging from a few 
metres to hundreds of metres), propagation loss and delay are negligible and the 
primary source of link loss is assumed to be the optical switches. For a 
link traversing $n_s$ intermediate switches, the total link transmission 
efficiency is approximated as $\eta_{\rm link} \approx {\eta'_s} ^{n_s}$, where
$\eta'_s = 10^{-\eta_s/10}$ is the per-switch transmission efficiency
and $\eta_s$ is the per-switch loss in dB.} In 
Sec.~\ref{sec:DQC_network_model}, we have defined two key parameters 
characterising the links in the logical topology: time required to establish 
entanglement ($T_{j_1 j_2}$) and fidelity ($F_{j_1 j_2}$) of entanglement 
generation for remote gates. Here, we specify their values for the 4-pod 
fat-tree topology used in our simulations. This topology features three 
types of links, distinguished by the number of optical switches along the 
path, specifically, 1, 3, or 5 switches. We denote the corresponding 
latency values by $T_{\text{link}_1}$, $T_{\text{link}_3}$, and 
$T_{\text{link}_5}$, respectively. Assuming an elementary link latency of 
$T_{\text{el}}$ (corresponding to entanglement generation time between QPUs 
without loss from any optical switches), and noting that the entanglement 
generation rate is directly affected by link loss, the latency for each 
link type is calculated as:
\begin{align}
T_{\text{link}_{n_s}} = \frac{T_{\text{el}}}{{\eta'_s}^{n_s}},
\end{align}
where $n_s \in 
\{1,3,5\}$. In our simulations, 
unless otherwise stated, all optical switches are assumed to be of the 
same type with an identical loss of $0.5$ dB. For the fidelity parameter, 
we assume a slight fidelity degradation with increasing link distance and 
loss, as follows: $F_{\text{link}_1}=0.96$, $F_{\text{link}_3}=0.94$, 
and $F_{\text{link}_5}=0.92$.
\begin{figure}[tb]
	\centering
        \includegraphics[width=0.75\linewidth]{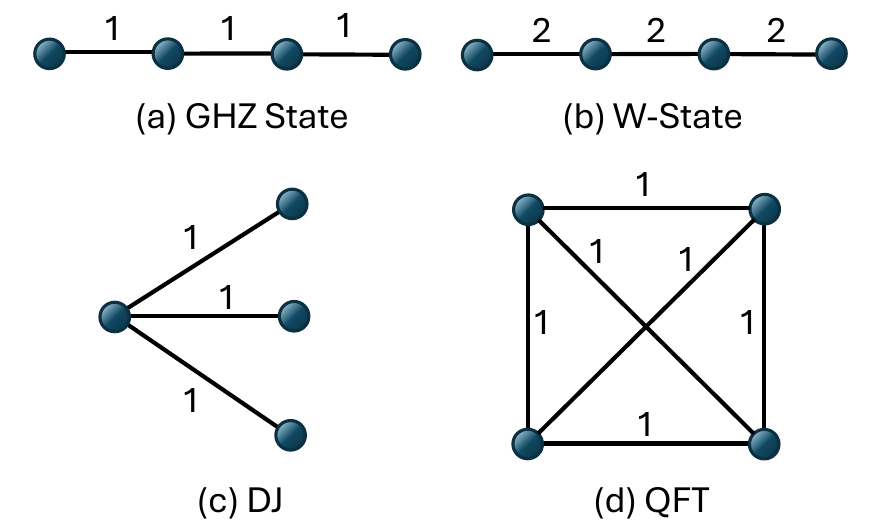}
        \vspace{-2mm}
	\caption{\small{Connectivity graphs for benchmark circuits ($w=4$). Edge weights indicate the number of two‑qubit gate occurrences.}} 
    \vspace{-3mm}
	\label{fig:QCirc_graph}  
\end{figure}
{
\subsubsection{Scheduling-Partitioning Configurations}

We evaluate the proposed algorithms under two categories of scheduling-partitioning configurations.

\noindent \textbf{(a) K-L-based scheduling-partitioning schemes:}
In the first category, we consider four scheduling strategies:
\begin{itemize}
    \item \textbf{Random Baseline (R-B)}: Circuits are randomly assigned to available QPUs subject to capacity feasibility. 
    \item \textbf{Capacity-Aware Baseline (CA-B)}: This baseline formulates QPU allocation as a classical knapsack problem~\cite{KelPfePis04}, accounting for QPU capacity constraints and 
circuit qubit demand while ignoring quantum-specific costs such as circuit 
connectivity, link fidelity, and decoherence. Assuming all-to-all network 
connectivity, it minimises the number of allocated QPUs subject to capacity 
and $K_{\rm max}$ constraints. The problem is solved to optimality using 
the same Gurobi solver as our proposed methods, providing a fair basis for 
comparison.
    \item \textbf{Single-QCirc}: The scheduler presented in Sec.\ref{sec:single_scheduler}, which optimises QPU allocation for each circuit individually.
    \item \textbf{Batch-QCirc}: The scheduler presented in Sec.\ref{sec:batch_scheduler}, which jointly optimises QPU allocation and partition counts for a batch of circuits. 
\end{itemize}

For each QCirc, once QPU allocation is determined by the scheduling policy, circuit partitioning is performed using a $k$-way graph partitioning approach based on iterative applications of the K-L algorithm. In what follows these three scheduling-partitioning settings are referred to as \textbf{R-B+K-L}, \textbf{CA-B+K-L}, \textbf{Single+K-L}, and \textbf{Batch+K-L}.

\noindent \textbf{(b) Pytket-dqc-based scheduling–partitioning schemes:}

In the second category, we adopt the pytket-dqc compiler framework 
\cite{andres2024distributing}, which supports circuit placement and 
partitioning over heterogeneous quantum networks using advanced compilation 
techniques. Unlike our formulation, pytket-dqc optimises individual 
circuits and does not explicitly consider multi-circuit scheduling under 
resource contention.

We consider two pytket-dqc distributors: \textit{PartitioningAnnealing} 
(PA), which employs simulated annealing on the circuit hypergraph 
representation, and \textit{PartitioningHeterogeneous} (PH), which uses 
the KaHyPar hypergraph partitioner. In all configurations, the selected distributor determines qubit placement and circuit partitioning, and the resulting allocation is translated into a distributed circuit using pytket-dqc’s standard compilation workflow.

Based on this integration, we evaluate six scheduling–partitioning configurations:

\begin{itemize}
    \item \textbf{Pytket-PA:} Each QCirc is processed sequentially using pytket-dqc’s native placement and partitioning mechanisms with the PA distributor.

    \item \textbf{Pytket-PH:} Same as above, but using the PH distributor.

    \item \textbf{Single+PA:} QPU allocation and scheduling decisions are determined by the proposed Single-QCirc scheduler, while pytket-dqc performs partitioning restricted to the subset of QPUs allocated to each circuit using PA.

    \item \textbf{Single+PH:} Same as above, but using PH.

    \item \textbf{Batch+PA:} QPU allocation and scheduling decisions are determined by the proposed Batch-QCirc scheduler, after which pytket-dqc performs partitioning within the allocated QPU subset using PA.

    \item \textbf{Batch+PH:} Same as above, but using PH.
\end{itemize}

In the latter four configurations, the network graph supplied to pytket-dqc is restricted to the QPUs selected by the scheduler, thereby cleanly separating high-level scheduling decisions from compiler-level partitioning. The two Pytket-only configurations serve as state-of-the-art compiler baselines, enabling direct comparison between standalone compiler-driven distribution and the proposed approaches.
}
\subsubsection{Benchmark QCircs and Workload Scenarios}

To establish the set of quantum circuits $\mathcal{QC}$, we use benchmark 
circuits from the Munich Quantum Toolkit~\cite{quetschlich2023mqtbench}: 
Quantum Fourier Transform (QFT), Deutsch--Jozsa (DJ), W-state, and GHZ-state 
circuits. Fig.~\ref{fig:QCirc_graph} shows connectivity-graph representations 
for $w=4$ qubits. From this figure, it can be inferred that GHZ is sparsely 
connected, W-state exhibits low-to-moderate connectivity, DJ circuits exhibit 
an intermediate level of connectivity, and QFT circuits, modelled with a 
complete graph, are highly connected.

A set $\mathcal{QC}$ of size $M$ includes all four circuit types, with each 
type contributing $25\%$ of the total. For circuit ${\rm QCirc}_m$, the number 
of qubits $w_m$ is randomly selected from a type-specific range. Two workload 
scenarios are considered to evaluate performance under increasing qubit demand. In \textit{Scenario~1 (Sc.1)}, the required qubit 
ranges are $R_{\text{GHZ}} = R_{\text{WState}} = [18,26]$, 
$R_{\text{DJ}} = [14,22]$, and $R_{\text{QFT}} = [10,18]$, where smaller 
ranges are assigned to QFT and DJ to account for their higher connectivity 
requirements. In \textit{Scenario~2 (Sc.2)}, the qubit demand is increased by 
shifting each range upward by 4, yielding 
$R_{\text{GHZ}} = R_{\text{WState}} = [22,30]$, $R_{\text{DJ}} = [18,26]$, 
and $R_{\text{QFT}} = [14,22]$. Simulations are conducted for $M \in \{12, 20, 28, 36\}$, and for each value 
of $M$, the experiment is repeated over 10 independently generated 
$\mathcal{QC}$ sets with results averaged across runs.

\begin{figure*}[!t]
    \centering
    \includegraphics[width=0.95\textwidth]{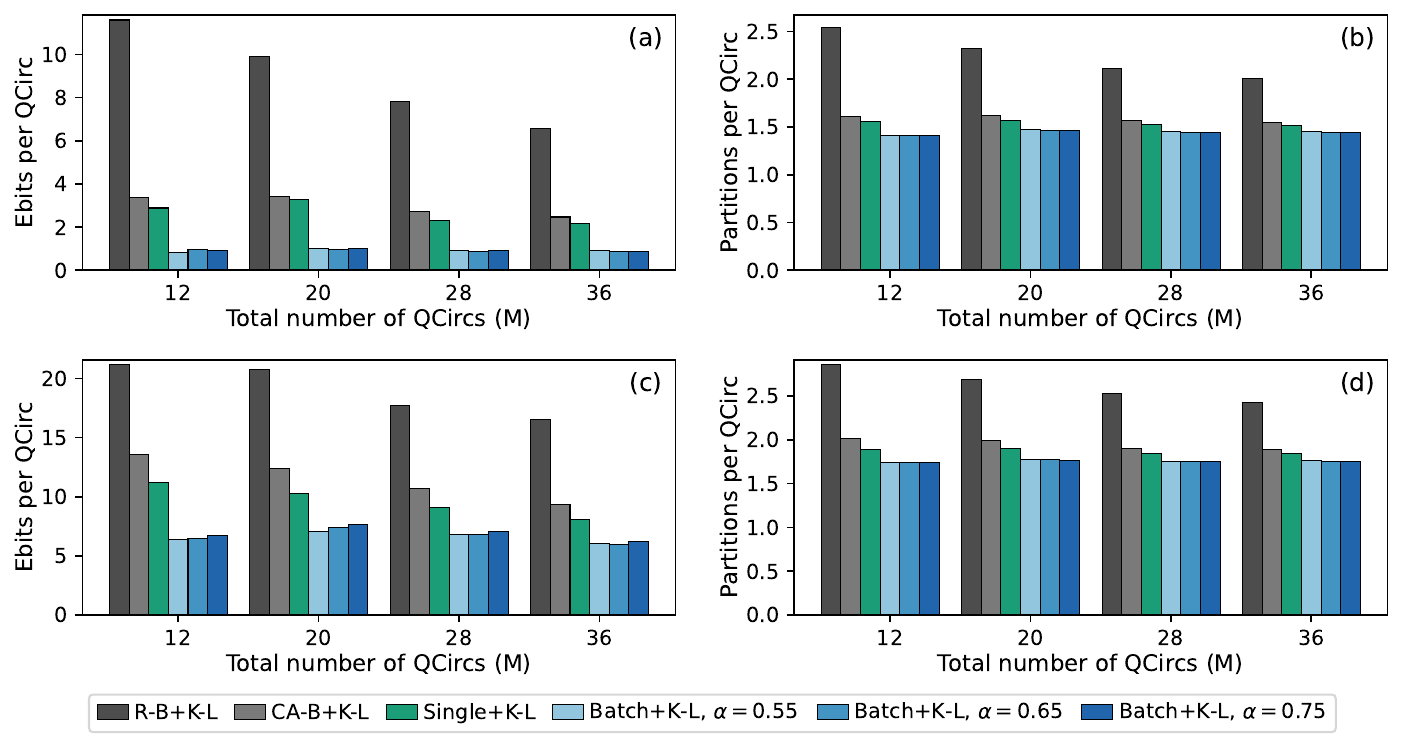}
    \vspace{-3mm}
    \caption{\small{Average number of ebits and circuit partitions divided by the total number of QCircs $M$, for the proposed and baseline scheduling-partitioning schemes. (a)-(b) Qubit-range scheme Sc.1, (c)-(d) Qubit-range scheme Sc.2.
}}
    \label{fig:total_Nrg}
    \vspace{-3mm}
\end{figure*}

\begin{figure*}[!t]
    \centering
    \includegraphics[width=0.95\textwidth]{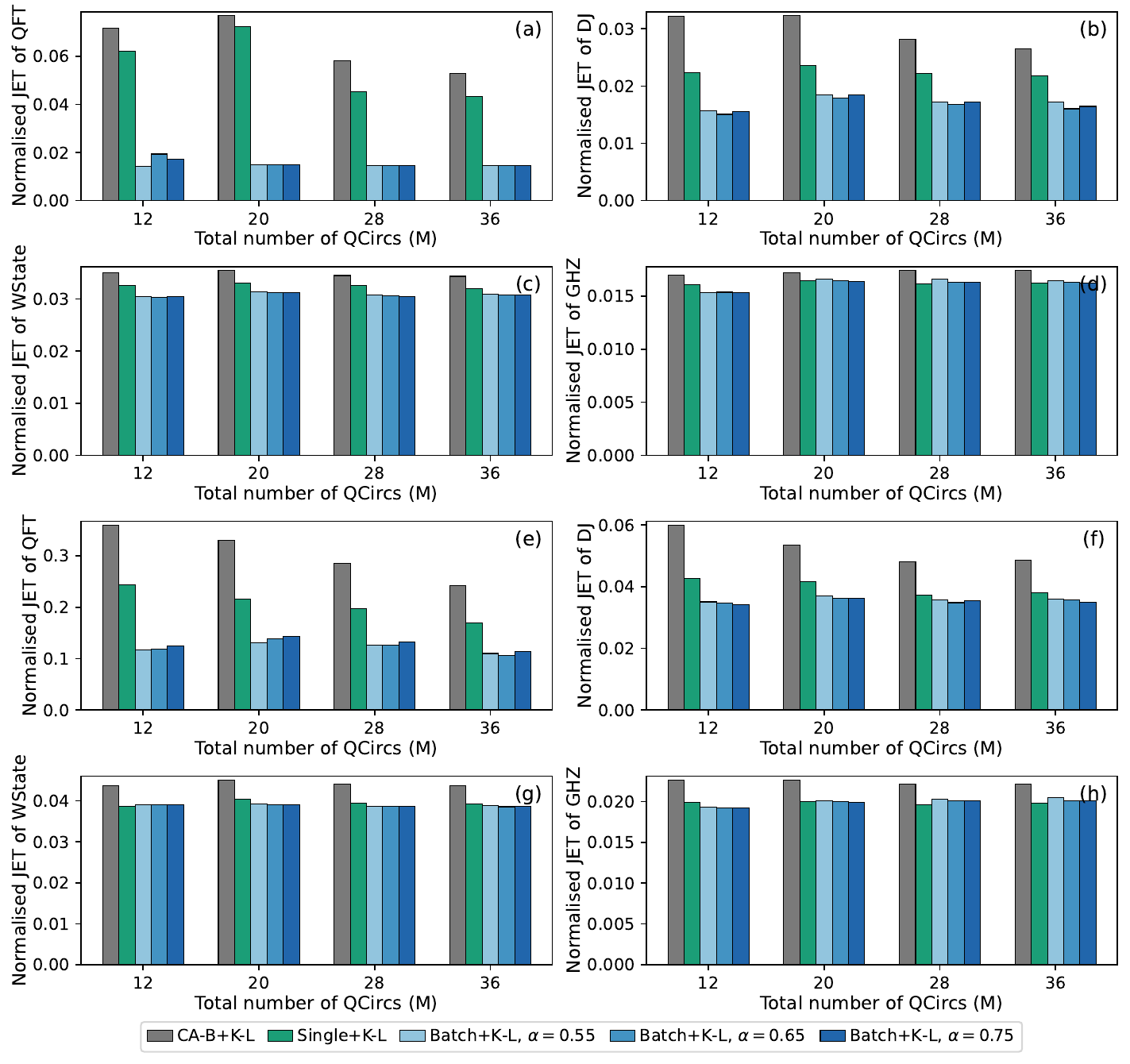}
    \vspace{-3mm}
\caption{\small{Average Job Execution Time (JET) (normalised by $T_{\text{dec}}$) for the proposed and baseline scheduling-partitioning schemes across different circuit types. (a)--(d) Qubit-range scheme Sc.1,  (e)--(h): Qubit-range scheme Sc.2.}}
    \label{fig:JET}
    \vspace{-3mm}
\end{figure*}

\subsubsection{Figures of merit}

To evaluate the performance of the proposed scheduling methods, we consider several key figures of merit. All results are averaged over repeated simulation runs. 

\begin{itemize}
    \item {\textbf{Number of ebits per QCirc}: the total number of ebits across all \(M\) QCircs, divided by \(M\), representing the quantum communication overhead. }
    \item \textbf{Number of circuit partitions per QCirc}: the total number of circuit partitions across all \(M\) QCircs, divided by \(M\).
    \item \textbf{Normalised Job Execution Time (JET) for each QCirc type}: Average execution time for a given QCirc type, computed over all its occurrences within $\mathcal{QC}$ and normalised by \(T_{\rm dec}\), i.e., JET/\(T_{\rm dec}\).
    \item \textbf{Number of ebits for each QCirc type}: Average number of ebits for a given QCirc type, computed over all its occurrences within $\mathcal{QC}$.
    \item \textbf{Normalised Makespan}: the total time required to complete all jobs (i.e., the execution of all \(M\) QCircs), normalised by \(T_{\rm dec}\), for consistency with normalised JET and to ensure hardware platform independence.
    \item \textbf{Normalised Throughput}: defined as the ratio of the number of QCircs (\(M\)) to Normalised Makespan. This indicates how many QCircs can be executed per unit time.
\end{itemize}

The normalised JET for a given QCirc is computed based on a layered circuit model, where the circuit is represented as a sequence of layers, and each layer comprises quantum gates that can be executed in parallel. First, for a given QCirc, remote entanglement generation is assumed to occur sequentially; parallel entanglement generation is not considered, even when multiple remote gates appear within the same circuit layer. Second, all local gates are assumed to have identical execution durations. Under these assumptions, we iterate over all circuit layers and determine whether each layer contains remote gates or only local gates. Let $N_{\rm LL}$ denote the number of layers containing exclusively local gates. The normalised JET is then computed as follows:
\begin{equation}
    \frac{\text{JET}}{T_{\rm dec}}=N_{\rm LL} \frac{T_{\rm local}}{T_{\rm dec}}+ \sum_{n_s \in \{1,3,5\}}{N^{(e)}_{\text{link}_{n_s}}\frac{T_{\text{link}_{n_s}}}{T_{\rm dec}}}
    \label{JET_calc}
\end{equation}
where $N^{(e)}_{\text{link}_{n_s}}$ denotes the number of ebits in the QCirc under consideration that are generated over links containing $n_s$ optical switches. In all simulations, we assume $T_{\rm local}/T_{\rm dec} = 5 \times 10^{-4}$ and $T_{\rm el}/T_{\rm dec} = 0.005$.

\subsection{Regression-Based Estimation of $\nu_{mk}$}
\begin{table}
\centering
\caption{\small{Linear regression results for our benchmark circuits.}}
\begin{tabular}{ccccccc}
\hline
&$\hat{\chi}_{0,k}$&$\hat{\chi}_{1,k}$&$\hat{\chi}_{2,k}$&$\hat{\chi}_{3,k}$&$R^2_{\rm test}$&${\rm RMSE}_{\rm test}$\\
\hline
$k=2$&0.0272&0.4345&0.0163&0.0434&0.995&0.0175\\
$k=3$&0.1185&0.4808&0.0534&0.0802&0.986&0.0369\\
$k=4$&0.1887&0.4585&0.081&0.1235&0.973&0.055\\
$k=5$&0.2842&0.3836&0.119&0.162&0.953&0.075\\
$k=6$&0.368&0.3&0.152&0.0203&0.925&0.096\\
\hline
\end{tabular}
\vspace{-5mm}
\label{tab:ridge_results}
\end{table}

This subsection presents the implementation details and performance evaluation of the linear regression model used to estimate the partitioning cost coefficient $\nu_{mk}$. The model is implemented using the \texttt{scikit-learn} library, while graph-related computations are performed with \texttt{NetworkX}.

{
The training dataset consists of all four circuit types with qubit counts $w_m \in [10,30]$ (84 circuits). For each circuit, the connectivity features $\gamma_m$, $\lambda_{2,m}$, and $\sigma_m/g_m$ are extracted from its graph representation. The regression target (see (\ref{cutsize})) is obtained by computing the balanced $k$-way graph cut size via iterative applications of the Kernighan–Lin (K–L) algorithm. The estimated partitioning cost $\nu_{mk}$ is then computed using the linear model defined in (\ref{LR}).

For evaluation, a separate test set comprising all four circuit types with $w_m \in [31,40]$ (40 circuits) is used. The learned coefficients, along with the test-set $R^2$ scores and root mean square error (RMSE), are reported in Table~\ref{tab:ridge_results}. The results demonstrate strong predictive performance across different partition counts. Among the features, $\lambda_{2,m}$ has the strongest influence, whereas the coefficients associated with $\gamma_m$ and $\sigma_m/g_m$ are generally smaller. As the partition count $k$ increases, the influence of $\gamma_m$ and $\sigma_m/g_m$ becomes more pronounced. Although prediction accuracy slightly decreases with larger $k$, the $R^2$ score remains above 0.9 in all cases. Since the test set contains circuits with qubit counts strictly outside the training range, the consistently high $R^2$ scores suggest that the model generalises well to unseen circuit sizes.

Note that generalisation to circuit types not present in the training set is not evaluated here, as the model is intended for use with the specific circuit types considered in this work. A more comprehensive study involving diverse circuit types and more sophisticated regression models is left for future work.
}
\subsection{Results for K-L-based scheduling-partitioning schemes }

In this subsection, we present the simulation results for the K-L-based scheduling-partitioning schemes. Fig.~\ref{fig:total_Nrg} shows the average number of ebits and circuit partitions per QCirc for the proposed and baseline methods. The results are reported for the qubit-range schemes Sc.1 and Sc.2 used in our simulations. For the Batch-QCirc scheduling method, several values of the parameter \(\alpha\) (the fraction threshold of the total qubits available across all QPUs before executing the next batch) are considered. As shown, both the Batch-QCirc and Single-QCirc methods outperform the baseline schemes. Moreover, the Batch-QCirc approach achieves better performance than the Single-QCirc method, albeit at the cost of a more complex MILP optimisation process. Since the number of ebits directly reflects the demand for inter-QPU communication resources, these results demonstrate that both proposed methods, and particularly the Batch-QCirc assignment, effectively reduce inter-QPU communication overhead.

{It can also be observed that the Random Baseline (R-B) consistently results in substantially higher ebit and partition counts than Capacity-Aware Baseline (CA-B) and the proposed methods. While it is included in Fig.~\ref{fig:total_Nrg} for completeness, it is excluded from the subsequent figures to maintain clarity and appropriate scaling.
\begin{figure*}[htb]
    \centering
    \includegraphics[width=0.95\textwidth]{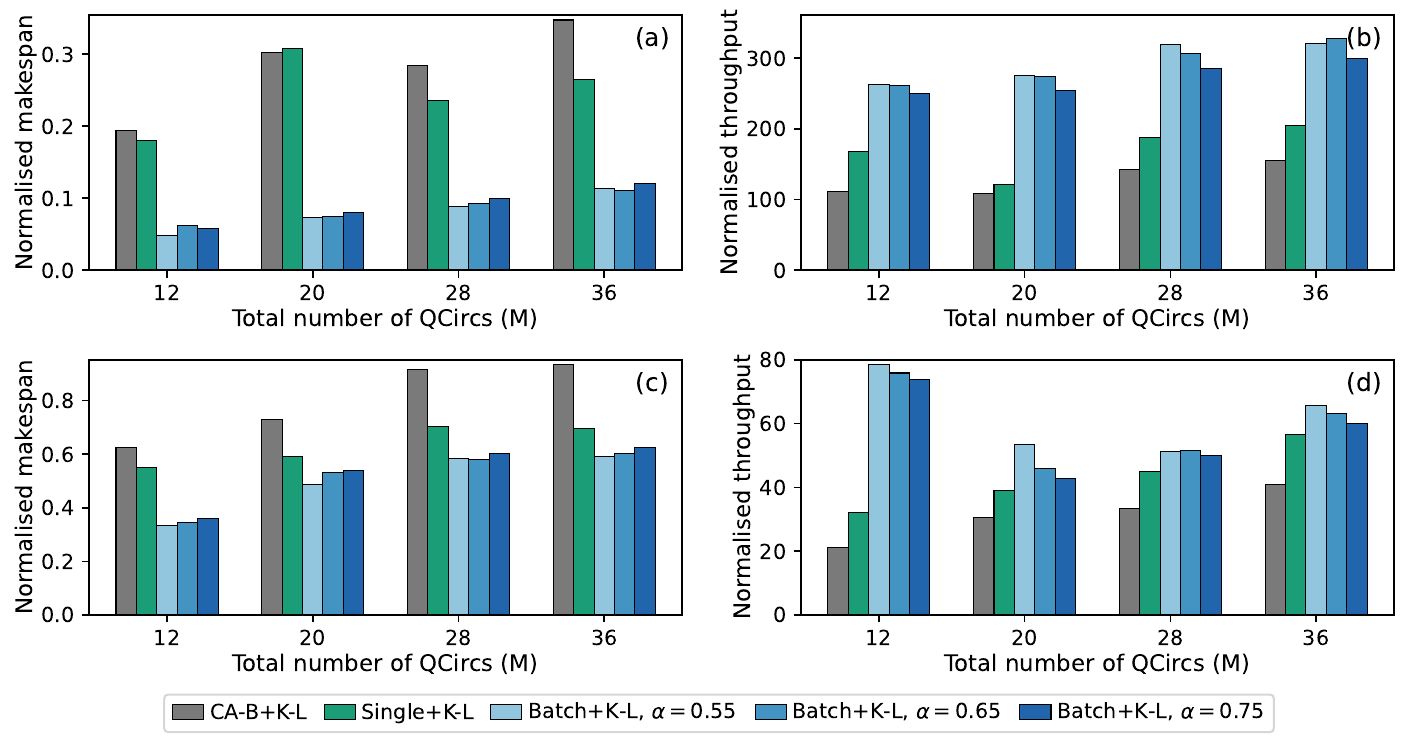}
    \vspace{-3mm}
    \caption{\small{Average makespan and throughput (normalised by $T_{\text{dec}}$) for the proposed and baseline scheduling-partitioning schemes. (a)-(b) Qubit–range scheme Sc.1, (c)-(d) Qubit–range scheme Sc.2.}}
    \label{fig:makespan}
    \vspace{-3mm}
\end{figure*}

\begin{figure}[!h]
    \centering
    \includegraphics[width=0.45\textwidth]{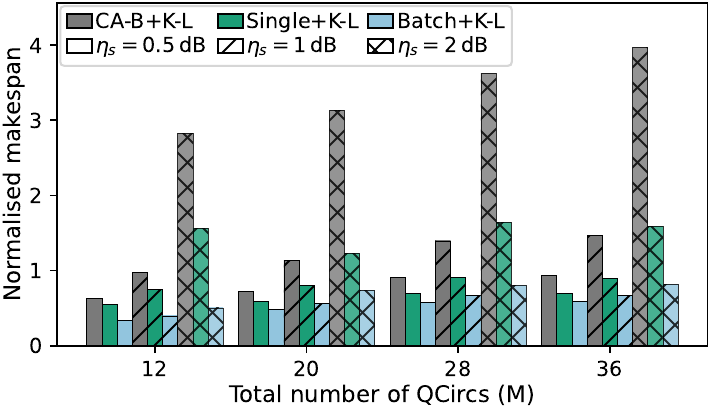}
    \caption{\small{Average makespan (normalised by $T_{\text{dec}}$) for the proposed and baseline scheduling-partitioning schemes under varying optical switch loss, assuming the qubit-range scheme Sc.2 and $\alpha=0.55$. }}
    \label{fig:makespan_switch}
    \vspace{-3mm}
\end{figure}


Next, we evaluate the normalised Job Execution Time (JET), computed individually for each circuit type. As shown in Fig.~\ref{fig:JET}, the proposed methods consistently outperform CA-B, achieving reduced execution times across all QCirc types. The Batch-QCirc scheduling approach further improves JET in most cases compared to the Single-QCirc method. In particular, Batch-QCirc significantly reduces the execution time for circuits with highly connected structures, such as the QFT circuits considered in our simulations, under both Sc.1 and Sc.2. In the joint-optimisation–based Batch-QCirc scheduling framework, QPUs with greater capacity tend to be prioritised for circuits with higher connectivity, thereby improving fidelity and reducing inter-QPU communication. However, this may limit optimisation opportunities for less connected circuits, such as GHZ. This trade-off is evident in both Sc.1 and Sc.2, where the Single-QCirc approach occasionally achieves lower JET for GHZ circuits. Such interdependence highlights the system-level nature of Batch-QCirc scheduling, where resource trade-offs naturally arise.}

Next, we examine the normalised makespan and throughput, which are key metrics for evaluating scheduling efficiency and resource utilisation. The results are shown in Fig.~\ref{fig:makespan}. The proposed methods improve both metrics, with the Batch-QCirc approach generally outperforming the Single-QCirc method. These results suggest that reductions in the required number of ebits can contribute to shorter makespan and higher throughput under the evaluated settings. We also compare Batch-QCirc scheduling with varying values of \(\alpha\), examining all the results presented in this section. In terms of makespan and throughput, the Batch-QCirc scheduling with \(\alpha=0.75\) consistently results in a slightly worse performance compared to other values. Since the parameter \(\alpha\) determines when the algorithm proceeds to the next scheduling cycle, it plays a crucial role in affecting these evaluated metrics. For the results of average number of ebits and JET, similarly, \(\alpha=0.75\) does not show meaningful improvements over \(\alpha=0.55\) or \(\alpha=0.65\); In fact, the latter two often outperform \(\alpha=0.75\). These results suggest that increasing waiting time for scheduling the next batch with a higher \(\alpha\) may lead to unnecessary resource waste without performance gains.

To investigate the impact of optical switch on the performance of the proposed DQC scheme, we increase the optical switch loss, which consequently raises the average entanglement generation times, $T_{\text{link}_{1}}$, $T_{\text{link}_{3}}$, and $T_{\text{link}_{5}}$. {Fig.~\ref{fig:makespan_switch} presents the makespan results for the CA-B, Single-QCirc and Batch-QCirc ($\alpha = 0.55$) methods under varying optical switch losses, $\eta_{s}$ = 0.5 dB, 1 dB, and 2 dB, assuming the qubit-range scheme Sc.2.} The results demonstrate that as $\eta_{s}$ increases, the relative advantage of Batch-QCirc becomes more pronounced. For instance, for $M=36$, Batch-QCirc reduces the makespan by $14.7\%$ compared to Single-QCirc at $\eta_s=0.5~\mathrm{dB}$, with the improvement increasing to $28.3\%$ and $43.8\%$ at $\eta_s=1~\mathrm{dB}$ and $\eta_s=2~\mathrm{dB}$, respectively. These results highlight the particular promise of the Batch-QCirc approach in topologies with heterogeneous link qualities.

\begin{figure*}[htb]
    \centering
    \includegraphics[width=0.95\textwidth]{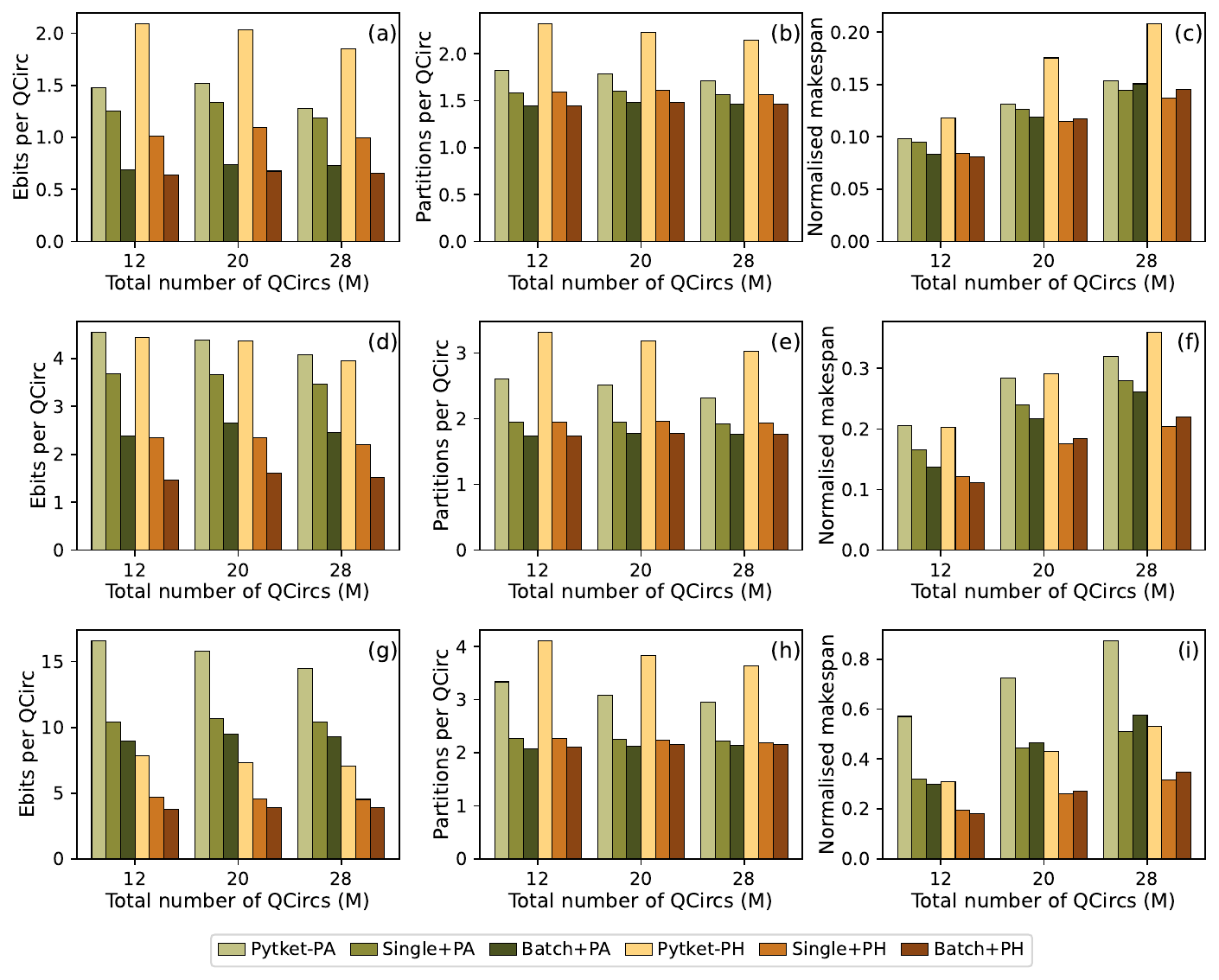}
    \vspace{-3mm}
    \caption{\small{Average number of ebits and circuit partitions divided by the total number of QCircs M, and Average makespan (normalised by $T_{\text{dec}}$) for the proposed and baseline Pytket-dqc-based scheduling-partitioning schemes. (a)-(c) Qubit–range scheme Sc.1, (d)-(f) Qubit–range scheme Sc.2, (g)-(i) Qubit–range scheme Sc.3.  }}
    \label{fig:pytket}
    \vspace{-3mm}
\end{figure*}

{
\subsection{Results for pytket-dqc-based scheduling-partitioning schemes}

In this subsection, we present the simulation results for the {pytket-dqc}-based scheduling–partitioning schemes. In addition to Sc.1 and Sc.2, we introduce an additional scenario, Sc.3, representing a partition-intensive regime with qubit ranges $R_{\text{QFT}} = R_{\text{DJ}} = [22,30]$ and $R_{\text{GHZ}} = R_{\text{WState}} = [24,32]$. In Sc.3, all circuits exceed the maximum single-QPU capacity of 20 qubits, ensuring that distributed execution with inter-QPU communication is unavoidable. This scenario is therefore designed to stress-test the combined scheduling–partitioning pipeline under fully distributed execution conditions. In our simulations, we consider $M \in \{12, 20, 28\}$.

Figure~\ref{fig:pytket} illustrates representative results, including average ebits per QCirc, average partition count per QCirc, and normalised makespan, for all six {pytket-dqc}-based scheduling–partitioning schemes under Sc.1, Sc.2, and Sc.3. The results demonstrate that both proposed methods consistently reduce ebit consumption and partition counts compared to the baselines (Pytket-PA and Pytket-PH). Comparing Single-QCirc and Batch-QCirc within these schemes, Batch-QCirc achieves lower ebit consumption and partition counts across all scenarios. However, Single-QCirc yields slightly shorter makespan for $M=20$ and $M=28$. This highlights an important system-level trade-off: while Batch-QCirc reduces communication overhead and JET through joint optimisation across circuits, it introduces scheduling latency governed by $\alpha$, whereas Single-QCirc performs scheduling immediately upon circuit arrival.

\begin{table}[htbp]
\centering
\caption{\small{Average number of ebits per QCirc type, assuming $M=28$ and the 
PH distributor.}}
\vspace{-2mm}
\footnotesize 
\setlength\tabcolsep{2pt} 
\renewcommand{\arraystretch}{1.1} 
\begin{tabular}{c ccc ccc ccc}
\hline
& \multicolumn{3}{c}{PH} 
& \multicolumn{3}{c}{Single+PH} 
& \multicolumn{3}{c}{Batch+PH} \\
\cline{2-10}
& Sc.1 & Sc.2 &Sc.3& Sc.1 & Sc.2 &Sc.3& Sc.1 & Sc.2&Sc.3\\
\hline
QFT & 1.375 & 6.01 & 17.2 & 1.04 & 4.48 & 13.35 & 0 & 2.23 & 10.53  \\
DJ & 0.96 & 1.93 & 2.82 & 0.47 & 0.88 & 1.15 & 0.26& 0.67 & 1.01 \\
W-State &3.39 & 5.15 & 5.4 & 1.68 & 2.25 & 2.39 & 1.52 & 2.1& 2.67  \\
GHZ & 1.67 & 2.7 & 2.75 & 0.77 & 1.15& 1.19 & 0.82 & 1.09 & 1.27  \\
\hline
\end{tabular}
\vspace{-2mm}
\label{table:Nrg_type}
\end{table}

To further examine the behaviour of the proposed methods, Table~\ref{table:Nrg_type} reports the average number of ebits for each of the four circuit types individually in the case $M=28$ using the PH distributor. The results show that both proposed approaches reduce ebit consumption compared to the baseline, with Batch-QCirc consistently achieving the lowest values. This indicates that, under the evaluated settings, the observed makespan trade-off primarily arises from scheduling latency rather than increased communication overhead associated with specific circuit types.}
{

\subsection{Computational Complexity and MILP Solver Runtime}
In this subsection, we analyse the computational complexity of the proposed 
MILP formulations and discuss their scalability. Let $J$ denote the number of QPUs, $S$ the batch size, and 
$K = \max_m K_m^{\max}$. The linearisation introduces auxiliary variables 
over QPU pairs, where $P = \binom{J}{2}$. Formulation~1 (Batch-QCirc) 
contains $S(J + K + P + KP)$ binary variables and $1 + 2S + J + 3SP + 3SKP$ 
linear constraints, scaling as $\Theta(SKJ^2)$. Formulation~2 (Single-QCirc) 
contains $J + P$ binary variables and $2 + 3P$ constraints, scaling as 
$\Theta(J^2)$. For the simulation setting $J = 16$ and $K = 4$, 
Formulation~2 has 136 binary variables and 362 constraints, while 
Formulation~1 has $620S$ binary variables and $(17 + 1802S)$ constraints.

The expected batch size $S$ depends on the scheduling parameters $\alpha$ 
and $\beta$. Since Batch-QCirc is triggered when an $\alpha$-fraction of 
total qubit capacity becomes available, the expected batch size can be 
approximated as:
\begin{equation}
S_{\rm typ} \approx \left\lfloor 
\frac{\beta\,\alpha\,\sum_j N_j}{\mathbb{E}[w_m]} \right\rfloor.
\end{equation}
For $\sum_j N_j = 224$, $\beta = 0.85$, and $\mathbb{E}[w_m] \approx 23$ 
(Sc.2), this yields $S_{\rm typ} \approx 4$ for 
$\alpha = 0.55$. 

Table~\ref{tab:runtime} reports the average MILP solver runtime per call for Single-QCirc and Batch-QCirc ($\alpha = 0.55$) under Sc.1, Sc.2, and Sc.3. All experiments were conducted on a Ubuntu server equipped with an Intel Core Processor at 2.40\,GHz, 32\,GB RAM, running Gurobi~v11.0.0. As expected, Batch-QCirc exhibits significantly higher runtime than Single-QCirc due to its larger joint optimisation problem. In contrast, the runtime of Single-QCirc remains relatively stable across scenarios, consistent with its fixed per-circuit problem size. For Batch-QCirc, the runtime under Sc.1 is higher than under Sc.2, since the lower average qubit demand in Sc.1 leads to larger batch sizes and therefore larger MILP instances. Interestingly, Sc.3 yields the highest Batch-QCirc runtime despite having smaller batch sizes. This is because, in Sc.3, all circuits exceed single-QPU capacity, forcing partitioning for every circuit and thereby expanding the feasible decision space explored by the MILP solver. These results indicate that the runtime of Batch-QCirc is influenced not only by batch size but also by circuit demand characteristics and partitioning intensity. While the runtimes observed here are acceptable for the considered network scale, extending the approach to substantially larger systems may benefit from heuristic or decomposition-based strategies, which we identify as a direction for future work.

\begin{table}[!tb]
\centering
\caption{\small{Average MILP solver runtime (s).}}
\vspace{-2mm}
\footnotesize
\setlength\tabcolsep{4pt}
\renewcommand{\arraystretch}{1.1}
\begin{tabular}{c ccc}
\hline
& Sc.1 & Sc.2& Sc.3 \\
\hline
Single-QCirc          & 0.0062 & 0.0061 &0.0065\\
Batch-QCirc ($\alpha=0.55$) & 3.41 & 2.93& 4.14 \\
\hline
\end{tabular}
\vspace{-6mm}
\label{tab:runtime}
\end{table}
}

\section{Conclusion}
\label{sec:conclusion}

{In this work, we address the challenge of efficient resource management and QCirc scheduling for DQC. We propose two scheduling approaches: a dynamic Batch-QCirc scheduling method, which operates through sequential scheduling cycles with batch optimisation, and a Single-QCirc scheduling method that optimises QCirc allocation individually. To achieve optimal assignment of QCircs to QPUs, we develop MILP formulations that minimise errors arising from inter-QPU communication. These formulations explicitly account for network topology, link loss, QPU capacities, and QCirc structure, enabling communication- and resource-aware allocation decisions. More specifically, the Batch-QCirc approach enables system-level optimisation across multiple concurrent circuits, thereby reducing communication overhead and improving execution fidelity, especially in heterogeneous network settings. At the same time, scalability considerations, together with the observed trade-off between inter-QPU communication overhead and makespan, suggest that batch-based optimisation is especially advantageous in small- to medium-scale, communication-resource-constrained regimes. On the other hand, the Single-QCirc formulation offers a lightweight and responsive scheduling strategy more suitable for online scheduling. 


This work represents an important step toward quantum- and network-aware resource management in cloud-based quantum data centres. To extend this further, consideration of dynamic quantum link models, heterogeneous hardware characteristics, and larger-scale deployments presents as important directions for future work. In addition, incorporating priority- and deadline-aware scheduling policies, as well as exploring machine learning-based approaches for adaptive and scalable optimisation, could further enhance intelligent resource management in quantum cloud infrastructures.

}

\section*{Acknowledgment}
The authors would like to express their sincere gratitude to Reza Nejabati for his support and insightful technical discussions.

\bibliographystyle{IEEEtran}
\bibliography{Main_References}

\end{document}